\PassOptionsToPackage{svgnames,dvipsnames,table}{xcolor}

\documentclass[a4paper,fleqn]{cas-dc}

\usepackage[authoryear]{natbib}
\usepackage{lastpage}
\usepackage{placeins}  
\usepackage{lineno}
\usepackage{soul}

\def\tsc#1{\csdef{#1}{\textsc{\lowercase{#1}}\xspace}}
\tsc{WGM}
\tsc{QE}
\tsc{EP}
\tsc{PMS}
\tsc{BEC}
\tsc{DE}
\begin{document}
\let\WriteBookmarks\relax
\let\printorcid\relax
\def\floatpagepagefraction{1}
\def\textpagefraction{.001}

\shorttitle{Peculiar rainbows in Saturn’s E ring}

\shortauthors{Rubbrecht et~al.}

\title [mode = title]{Peculiar rainbows in Saturn’s E ring: Uncovering luminous bands near Enceladus}  

% \tnotemark[1,2]

% % Title footnote 1.
% % eg: \tnotetext[1]{Title footnote text}
% % \tnotetext[<tnote number>]{<tnote text>} 
% \tnotetext[1]{This document is the results of the research
%    project funded by the National Science Foundation.}

% \tnotetext[2]{The second title footnote which is a longer text matter
%    to fill through the whole text width and overflow into
%    another line in the footnotes area of the first page.}

\author[1]{Niels Rubbrecht}

% \author[1,3]{Niels Rubbrecht}[type=editor, style=chinese,
%                         auid=000,bioid=1,
%                         prefix=,
%                         role=Researcher,
%                         orcid=0000-0001-7511-2910]

% \ead[url]{www.cvr.cc, cvr@sayahna.org}

% \credit{Conceptualization of this study, Methodology, Software}

\affiliation[1]{organization={Faculty of Aerospace Engineering, Delft University of Technology},
    addressline={Kluyverweg 1}, 
    city={Delft},
    % citysep={}, % Uncomment if no comma needed between city and postcode
    postcode={2629 HS}, 
    country={The Netherlands}}

% PI
\author[1]{St\'ephanie Cazaux}

% Email id of the 2nd author
\ead{s.m.cazaux@tudelft.nl}

\author%
[2]{Beno\^it Seignovert}
% \cormark[2]
% \fnmark[1,3]
% \ead{rishi@stmdocs.in}
% \ead[URL]{www.stmdocs.in}

% \affiliation[2]{
%     organization={Observatoire des Sciences de l'Univers Nantes Atlantique (Osuna)}, 
%     addressline={CNRS UAR-3281, Nantes Université, France}
% }
% \affiliation[2]{
%     organization={Observatoire des Sciences de l'Univers Nantes Atlantique (Osuna), CNRS UAR-3281, Nantes Université, France}
% }
\affiliation[2]{
    organization={Observatoire des Sciences de l'Univers Nantes Atlantique (Osuna)},
    addressline={CNRS UAR-3281, Nantes Université},
    country={France}
}

\author%
[1, 3]{Matthew Kenworthy}
\affiliation[3]{
    organization={Leiden Observatory, Leiden University},
    addressline={P.O. Box 9513},
    city={Leiden},
    postcode={2300 RA},
    country={The Netherlands}
}

% Fifth author
\author%
[4]{Nicholas Kutsop}

\affiliation[4]{
    organization={Department of Astronomy, Cornell University},
    addressline={404 Space Sciences Building},
    city={Ithaca},
    state={NY},
    postcode={14853},
    country={United States}
}

% sixth
\author[5]{St\'ephane Le\,Mou\'elic}
\affiliation[5]{
    organization={Laboratoire de Planétologie et Géosciences (LPG), CNRS UMR 6112, Nantes Université, Univ Angers, Le Mans Université},
    city={Nantes},
    postcode={44000},
    country={France}
}

\author[1]{J\'er\^ome Loicq}

\begin{abstract}
We report observations of stripe-like features in Enceladus' plumes captured simultaneously by Cassini’s VIMS-IR and ISS NAC instruments during flyby E17, with similar patterns seen in VIMS-IR data from flyby E13 and E19. These parallel stripes, inclined at approximately 16° to the ecliptic and 43° to Saturn’s ring plane, appear continuous across images when projected in the J2000 frame. A bright stripe, most visible at wavelengths around 5 \textmu m, acts as the zeroth-order diffraction peak of a reflection grating with an estimated groove spacing of 0.12–2.60 mm, while adjacent stripes are attributed to higher-order diffraction peaks. We suggest that this light-dispersing phenomenon originates from an inclined periodic structure within Saturn's E ring. This structure, constrained between
Saturn’s G ring and Rhea’s orbit, likely consists of fresh
ice particles supplied by Enceladus’ plumes.
\end{abstract}

% \includegraphics{figs/grabs.pdf}

% \begin{highlights}
% \item Research highlights item 1
% \item Research highlights item 2
% \item Research highlights item 3
% \end{highlights}

% Keywords
% Each keyword is separated by \sep
\begin{keywords}
Enceladus \sep Ices, IR spectroscopy \sep Infrared observations \sep Saturn \sep Rings \sep Satellites
\end{keywords}

\maketitle

\section{Introduction}

% background
\noindent Since reaching the Saturnian system in 2004, NASA’s Cassini spacecraft has revolutionised our understanding of Saturn, its rings, and its diverse moons \citep{Spilker2019Cassini-HuygensDiscovery}. Among Cassini’s most significant discoveries was the detection of active water-ice plumes erupting from Saturn’s icy moon, Enceladus \citep{Porco2006CassiniEnceladus}. Cassini's remote sensing instruments, including the Imaging Science Subsystem (ISS) and Visual and Infrared Mapping Spectrometer (VIMS), detected light scattered by the micrometre-sized ice particles ejected from fissures on the moon’s south polar region \citep{Ingersoll2020TimeChange}. These observations have proven essential for refining models of plume particle size distributions and their dynamics \citep{Hedman2009SpectralCassini-vims, Ingersoll2011TotalImages}. It became evident that these plumes represent the primary source of particles contributing to Saturn's diffuse E ring \citep{Schmidt2008SlowFractures, Kempf2008TheParticles}. 

The plumes and E ring provide a unique opportunity to investigate Enceladus' internal processes and evolutionary history. In-situ analysis of E ring particles with Cassini's Cosmic Dust Analyzer (CDA) revealed the presence of ice enriched in sodium salts \citep{Postberg2009SodiumEnceladus}. When combined with gravity field measurements \citep{Park2024TheEnceladus}, these findings support the existence of a global subsurface ocean beneath Enceladus' icy crust, from which the plumes originate. As a result, the plumes and E ring particles provide direct samples of this subsurface ocean. Subsequent research detected complex organic molecules in the icy grains of the plumes \citep{Postberg2018PlumeEnceladus, Choblet2022EnceladusExplorations}. These findings point to a potentially habitable environment beyond Earth, which makes Enceladus a prime target for future missions aimed at searching for extraterrestrial life \citep{Carr2013LifeSequencer, Porco2017CouldMissions}. This emphasis is further supported by ESA's Voyage 2050 program, which prioritises the exploration of icy moons, including Enceladus, as part of its long-term strategy for Solar System exploration \citep{Favata2021IntroducingProgramme}. Awaiting future missions, Cassini's extensive observations from the ISS and VIMS instruments continue to yield new scientific insights nearly 20 years after its initial deployment, e.g. \cite{Morello2024AEnceladus} and \cite{Denny2024ConstrainingSpectrometer}.

This study introduces a novel remote-sensing technique aimed at characterising particles within Enceladus' plumes by observing optical phenomena similar to those seen in the Earth’s atmosphere, such as rainbows and halos \citep{Lynch1991RainbowsFogbows, Tape2006AtmosphericX}. These phenomena result from light interactions with particles like water droplets or ice crystals, providing insights into particle properties through unique visual effects and specific occurrence conditions \citep{Laven2004SimulationSeries, Nussenzveig2012THEGLORY, Moilanen2022LightHalos}. While rare beyond Earth, such phenomena have been recorded elsewhere in the Solar System. For instance, the Venus Express orbiter captured a glory on the upper cloud layer of Venus, shedding new light on the material and size distribution of the cloud particles \citep{Markiewicz2014GloryAbsorber, Petrova2015TheVenus}. Similarly, the Mars Perseverance rover documented an ice particle halo created by hexagonal ice crystals within water-ice clouds, offering new insights into Martian cloud formation conditions \citep{Lemmon2022HexagonalRover}. 

Images of Enceladus' plumes captured by the ISS and VIMS instruments provide a unique opportunity to study similar optical phenomena. These observations depend on light scattered by particles within the optically thin plumes \citep{Ingersoll2020TimeChange}, under conditions akin to those that create rainbows and halos in the Earth's atmosphere. In the search for optical phenomena in Enceladus' plumes, we identified stripe-like features captured simultaneously by Cassini's ISS and VIMS instruments. This study aims to investigate these features in detail and link them to underlying physical processes.
 
% structure 
This paper is articulated as follows. Cassini VIMS and ISS observations of the stripes in Enceladus targeted flybys E13, E17, and E19 will be detailed in \autoref{sec:obs_E17}. These, stripe observations will be characterised in terms of spectral variations and material properties in \autoref{sec:character}. In \autoref{sec:hypo}, the potential origin of the stripes will be discussed.

\section{Stripes in Cassini ISS \& VIMS Observations}
\label{sec:obs_E17}

\noindent This section reports an overview of Cassini's observations of similar stripes in the ISS and VIMS datasets during its targeted flybys of Enceladus (E13 on 2010-12-21, E17 on 2012-03-27, and E19 on 2012-05-02). These measurements will then be compared and related to Cassini's observation geometry for the Saturnian system. 

% single image to entire sequence 
\subsection{ISS Images}

\noindent The Cassini Imaging Science Subsystem (ISS) cameras, detailed in \cite{Porco2004CASSINISATURN}, consist of two fixed focal length telescopes: the Narrow Angle Camera (NAC) and the Wide Angle Camera (WAC). The NAC has a field of view (FOV) of $0.35^\circ \times 0.35^\circ$ and a spectral range of 200–1100 nm, while the WAC offers a wider FOV of $3.5^\circ \times 3.5^\circ$ and a spectral range of 380–1100 nm. Both cameras use filter wheels to combine spectral filters for flexible imaging capabilities. This research solely focuses on NAC images which have an angular resolution of 6 \textmu rad/pixel. The images are retrieved from NASA’s PDS Ring-Moon Systems Node's \href{https://pds-rings.seti.org/search/}{OPUS search service} and calibrated using the CISSCAL 4.0 pipeline \citep{Knowles2020End-of-missionSubsystem} to convert raw data to $I/F$, which is a standardised measure of reflectance that equals unity for a Lambertian surface at normal incidence and emission. The Cassini ISS data is handled using the Integrated Software for Imagers and Spectrometers (ISIS) version 8.0.0 developed by the USGS Astrogeology Science Center \citep{Laura2023IntegratedSpectrometers}. 

\subsection{VIMS Hyperspectral Cubes}

\noindent The Cassini Visual and Infrared Mapping Spectrometer (VIMS), described in \cite{Brown2004TheInvestigation}, is composed of two instruments, VIMS-VIS and VIMS-IR, which observe visible and infrared wavelengths, respectively. This study focuses solely on the IR channel, which captures spectra at 256 wavelengths between 0.85 and 5.1 \textmu m. It has an average spectral resolution of 16 nm and an angular spatial resolution of 0.25 x 0.5 mrad covering a maximum viewing window of $64 \times 64$ spatial pixels. The resulting spectral data product is known as a "cube." VIMS-VIS data was not acquired for most of the observations used in this report and is therefore not used in the subsequent analysis. The calibrated VIMS-IR spectral cubes are directly downloaded from \hyperlink{vims.univ-nantes.fr}{vims.univ-nantes.fr}, which contains a collection of multispectral summary products created by \cite{LeMouelic2019TheMaps}. These cubes are handled using the PyVIMS (Version 1.0.4) Python package developed by \cite{Seignovert2023PyVIMS}.

\subsection{E17 Flyby}

\noindent Stripe features were first identified in observations during Cassini's Enceladus targeted flyby E17. This subsection discusses the stripe observations captured by ISS and VIMS, analyses them within the context of the flyby approach, and compares the findings from both instruments. 

\subsubsection{Initial ISS Observations}

\begin{figure}[h]
	\centering
		\includegraphics[scale=.35]{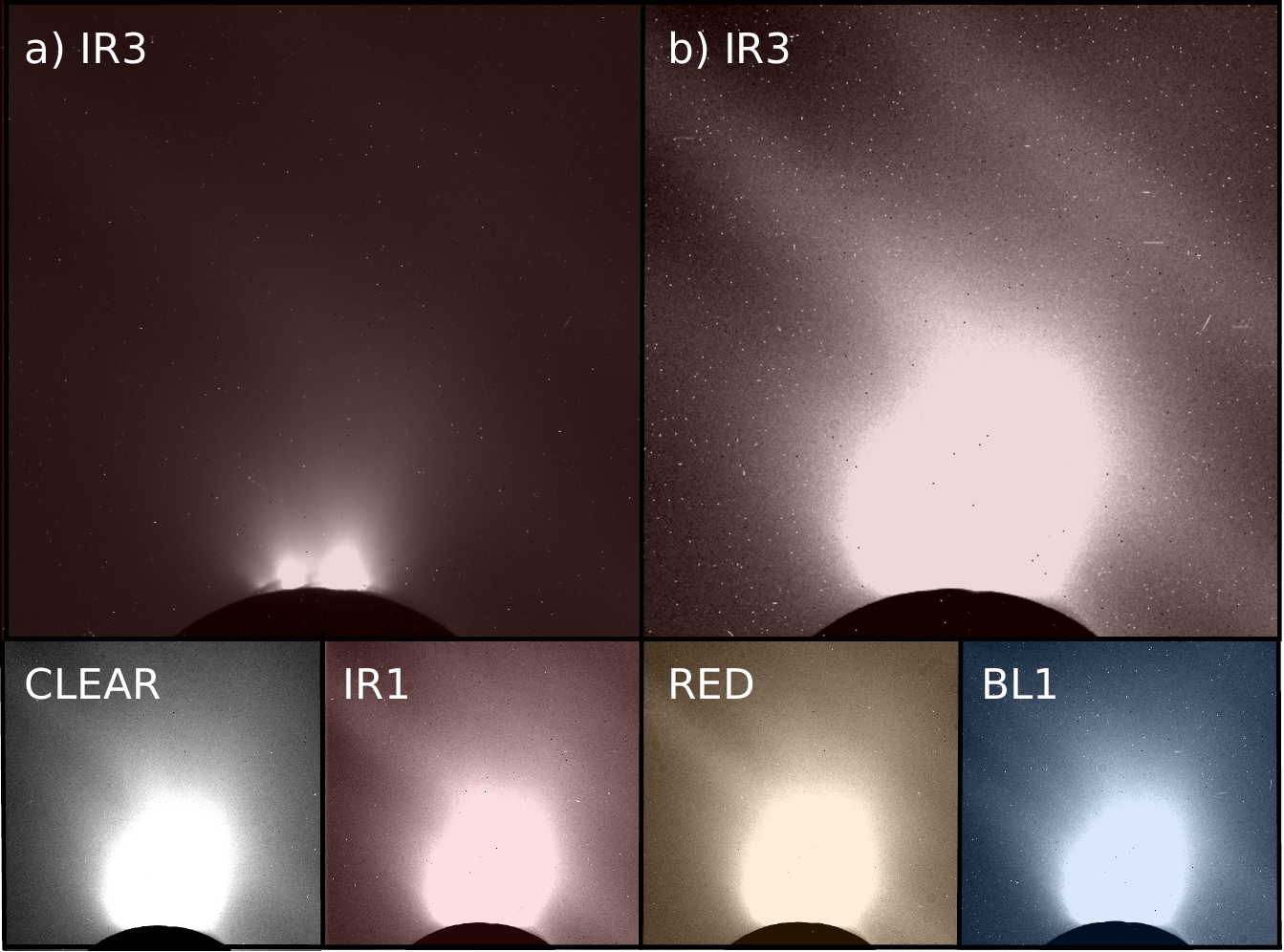}
	\caption{a) Cassini ISS NAC image N1711553432 of Enceladus' plumes with $I/F$ linear stretch of [$2.4 \times 10^{-4}$, $2.8 \times 10^{-3}$], and  b) [$4.7 \times 10^{-4}$, $7.6 \times 10^{-4}$]. The bottom row shows images N1711553312, N1711553502, N1711553572, and N1711553692, respectively, with the latter $I/F$ stretch.}
	\label{FIG:1}
\end{figure}

\noindent \autoref{FIG:1} a) shows ISS NAC image N$\mathrm{1711553432}$ of Enceladus' South pole and its plumes captured during targeted flyby E17. 
The image is shown over an $I/F$ range between $2.4 \times 10^{-4}$ and $2.8 \times 10^{-3}$. In this high phase (160$^\circ$) observation, the forward scattering of light hitting the plume particles reveals two bright jets emanating from Enceladus' body, which appears darker against the E ring background. In \autoref{FIG:1} b), the same image is overexposed with a linear $I/F$ stretch between $4.7 \times 10^{-4}$ and $7.6 \times 10^{-4}$. This range highlights three faint parallel stripes that appear oriented from the image's top left to the bottom right corner. The $I/F$ of the stripes is one order of magnitude lower than that of the brightest areas of the jets.

The bottom row of \autoref{FIG:1} displays four additional images, each captured within a maximum time difference of five minutes from N1711553432, resulting in similar observational geometries. Each image was taken using a different broad or medium-band filter, with the filter names indicated in the top left corners. In every image, similar parallel stripes can be seen. These stripes are thus visible across a spectral range covering both the visible and the near-infrared.  

\subsubsection{ISS Mosaic}

\noindent To investigate the extent of the stripe features found in \autoref{FIG:1}, the same linear stretch is applied to all plume images captured during Cassini's E17 approach. A total of ten images have been found to contain similar stripe features. The final 16 images out of the 60 captured during the E17 approach are mosaicked on the sky based on their right ascension ($RA$) and declination ($DEC$) in the J2000 reference frame. Here, the ISS NAC image corner coordinates are retrieved from "SPICE" (Spacecraft, Planet, Instrument, C-matrix, Events) system, designed by NASA's Navigation and Ancillary Information Facility (NAIF) \citep{Acton1996AncillaryFacility} and processed through SpiceyPy \citep{Annex2020SpiceyPy:Toolkit} for the image observation mid-times. 

\begin{figure}[h] 
	\centering
		\includegraphics[width=\linewidth]{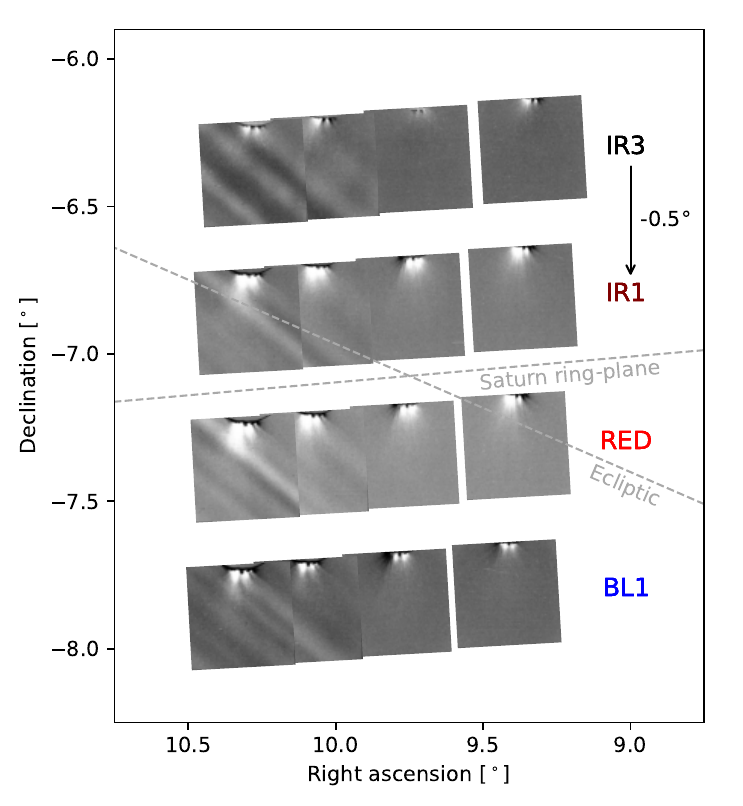}
	\caption{Cassini flyby E17 ISS NAC mosaic in terms of right ascension and declination in the J2000 reference frame for different medium-band filters. Images (\autoref{tbl1}) are subtracted by the nearest CLEAR filter image and have an $I/F$ range of $[-10^{-4}, 10^{-4}]$. The original mosaic is given on top; every following filter mosaic is shown with an additional declination of $-0.5^\circ$ for clarity.}
	\label{FIG:2}
\end{figure}

Projecting these coordinates results in the mosaic shown in \autoref{FIG:2}. During the approach, the ISS NAC cycles through a filter sequence in the order CLEAR--IR3--IR1--RED--BL1 in quick succession as shown in \autoref{tbl1}. After a short waiting period, the sequence repeats. In the mosaic, images captured at a later time are superimposed on the earlier ones. Therefore, images captured in a single sequence would appear overlapped due to the minor variation in observation geometry. To allow for comparison in \autoref{FIG:2}, the projections are separated for each filter by a DEC offset of $-0.5^\circ$. The original mosaic coordinates are displayed for images captured with the IR3 filter. The imaging sequence starts from the right side of the figure at $RA=9.2^\circ$ with the first IR3 image and ends with the last BL1 image at $RA=10.5^\circ$. The observations in \autoref{FIG:2} were further processed by subtracting the CLEAR images from the same sequence. This subtraction reduces the plumes' background brightness, enhancing the stripe features' visibility. The CLEAR and filtered images were subtracted without precise alignment, as the similar observational geometries within each filter sequence made this approach adequate.

The mosaic reveals that stripe features are visible only in the last eight images of the E17 approach, with the final four corresponding to those shown in \autoref{FIG:1}. Enceladus' South Pole is present in all 16 images in the top centre of each image window. The stripes appear near-continuous across images taken with the same filter. However, there is a slight misalignment where the last image is shifted by $DEC\approx 0.1^\circ$ compared to the second-to-last image. This misalignment could result from small pointing errors in the SPICE kernels or parallax. Nevertheless, the intensity variations remain consistent across the images, as seen in the order and intensity of the bright and dark regions in the IR1 images.

Comparing different filters reveals notable differences: the IR3 filter shows fewer, thicker stripes with high contrast, while the visible filters display more numerous, thinner, and closely spaced stripes. Stripe visibility also changes across filters, with some stripes appearing or disappearing depending on the filter used. Despite these variations, the stripes consistently run parallel throughout the mosaic and are inclined approximately 16$^\circ$ with respect to the ecliptic and $43^\circ$ inclined to Saturn's ring plane. An overview of the key observational parameters of the E17 images containing stripe features is shown in \autoref{tbl1}. Here, the phase is the mean phase angle of the pixels in the observation window.

\begin{table}[width=\linewidth,cols=4,pos=h]
\caption{E17 ISS NAC images containing stripes.}\label{tbl1}
\begin{tabular*}{\tblwidth}{@{} LLLL@{} }
\toprule
Image ID & Image Mid Time & Filter & Phase [$^\circ$]\\
\midrule
N1711552042	& 2012-03-27 14:15:56 & CLEAR & 160.22 \\
N1711552162 & 2012-03-27 14:17:24 & IR3   & 160.23  \\
N1711552232	& 2012-03-27 14:19:02 & IR1 &   160.25   \\
N1711552302 & 2012-03-27 14:20:13 & RED &   160.26  \\
N1711552422	& 2012-03-27 14:21:50 & BL1 &   160.27  \\
N1711553312 & 2012-03-27 14:37:06 & CLEAR &  160.39\\
N1711553432 & 2012-03-27 14:38:34 & IR3   & 160.40\\
N1711553502 & 2012-03-27 14:40:12 & IR1 & 160.41\\
N1711553572	& 2012-03-27 14:41:23 & RED &  160.42\\
N1711553692 & 2012-03-27 14:43:00 & BL1 & 160.43\\
\bottomrule
\end{tabular*}
\end{table}

\subsubsection{Initial VIMS Observations}

\begin{figure}[h]
	\centering
		\includegraphics[width=\linewidth]{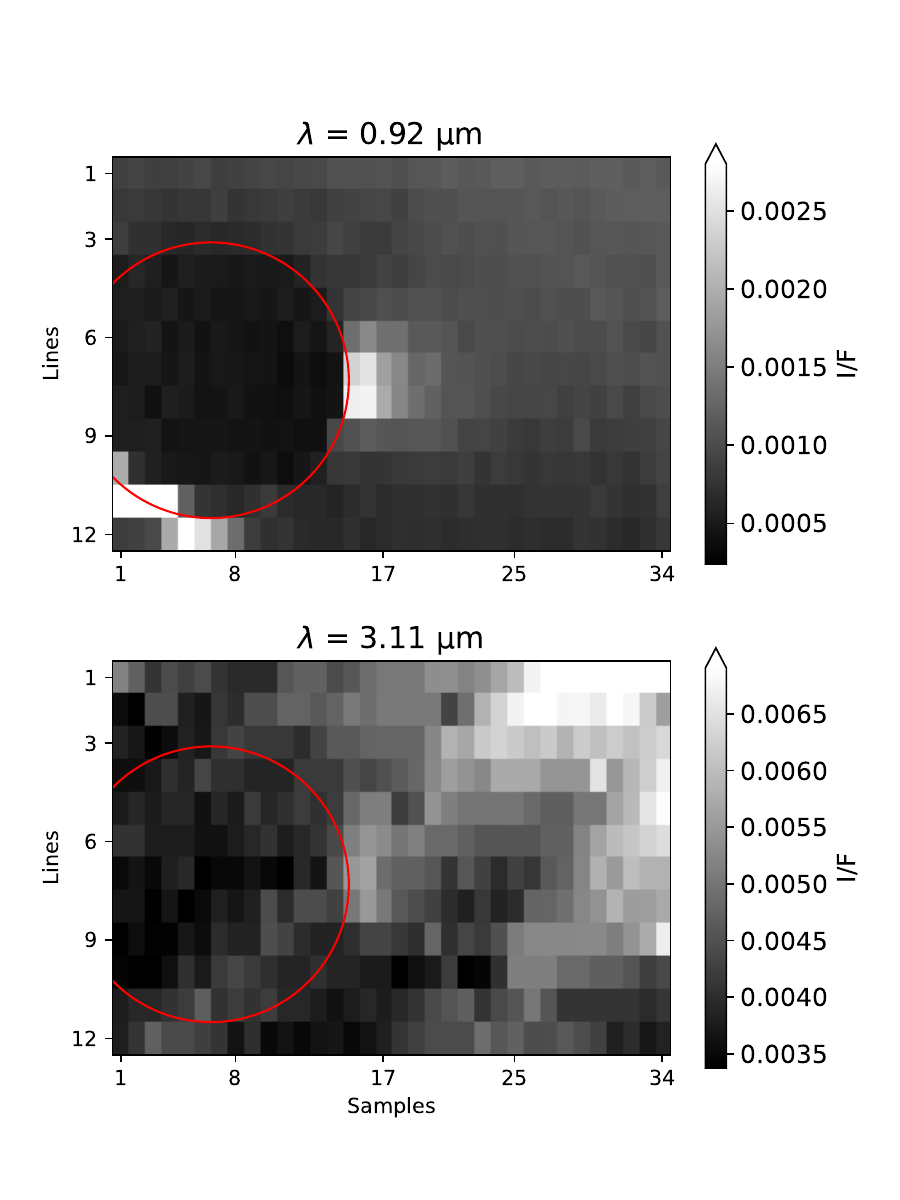}
	\caption{Unprojected VIMS cube 1711553290 with Image Mid Time $=$ 27/03/2012 14:39:09 showing $I/F$ for wavelengths 0.92 \textmu m and 3.11 \textmu m. The limb of Enceladus is highlighted in red for reference.}
	\label{FIG:3}
\end{figure}

\noindent During Cassini's E17 approach to Enceladus, simultaneous observations were made using the ISS and VIMS instruments. \autoref{FIG:3} shows VIMS cube $\mathrm{1711553290}$ which was captured with an image mid-time 35 seconds later than ISS NAC image N1711553432 from \autoref{FIG:1}. The top of the figure shows the $I/F$ values of the cube at wavelength $\lambda=0.92$ \textmu m, which is similar to the central wavelength of the ISS IR3 filter ($\lambda=0.93$ \textmu m) \citep{Knowles2018CassiniGuide}. For comparison, the same stretch of $I/F$ values is used as in \autoref{FIG:1} a). The dark body of Enceladus, circled in red, is seen against the faint E ring background. The bright pixels in the lower left corner of the image are the limb of Enceladus illuminated by the Sun. The bright pixels in the centre of the image are caused by the forward scattered light from the plumes. No stripes can be identified in this image, as was the case for \autoref{FIG:1} a). Considering an interval of the $I/F$ values similar to ISS shown in \autoref{FIG:1} b) leads to the saturation of most of the pixels in the image and thus was not reported. However, the same cube for $\lambda=3.11$ \textmu m is reported instead, highlighting a large diagonal stripe, which is brightest in the top right corner and reduces in intensity when moving to the bottom left corner of the image. Note that the stripe appears in front of the dark body of Enceladus. A second stripe can be observed to the right of the former. Both stripes appear approximately parallel to each other. In addition, the bright region in the corner is significantly brighter than the plume itself with an increase of $I/F\approx$ $10^{-3}$. 
   
\subsubsection{VIMS Mosaic}

\begin{figure*}[h] 
	\centering
		\includegraphics[width=\linewidth]{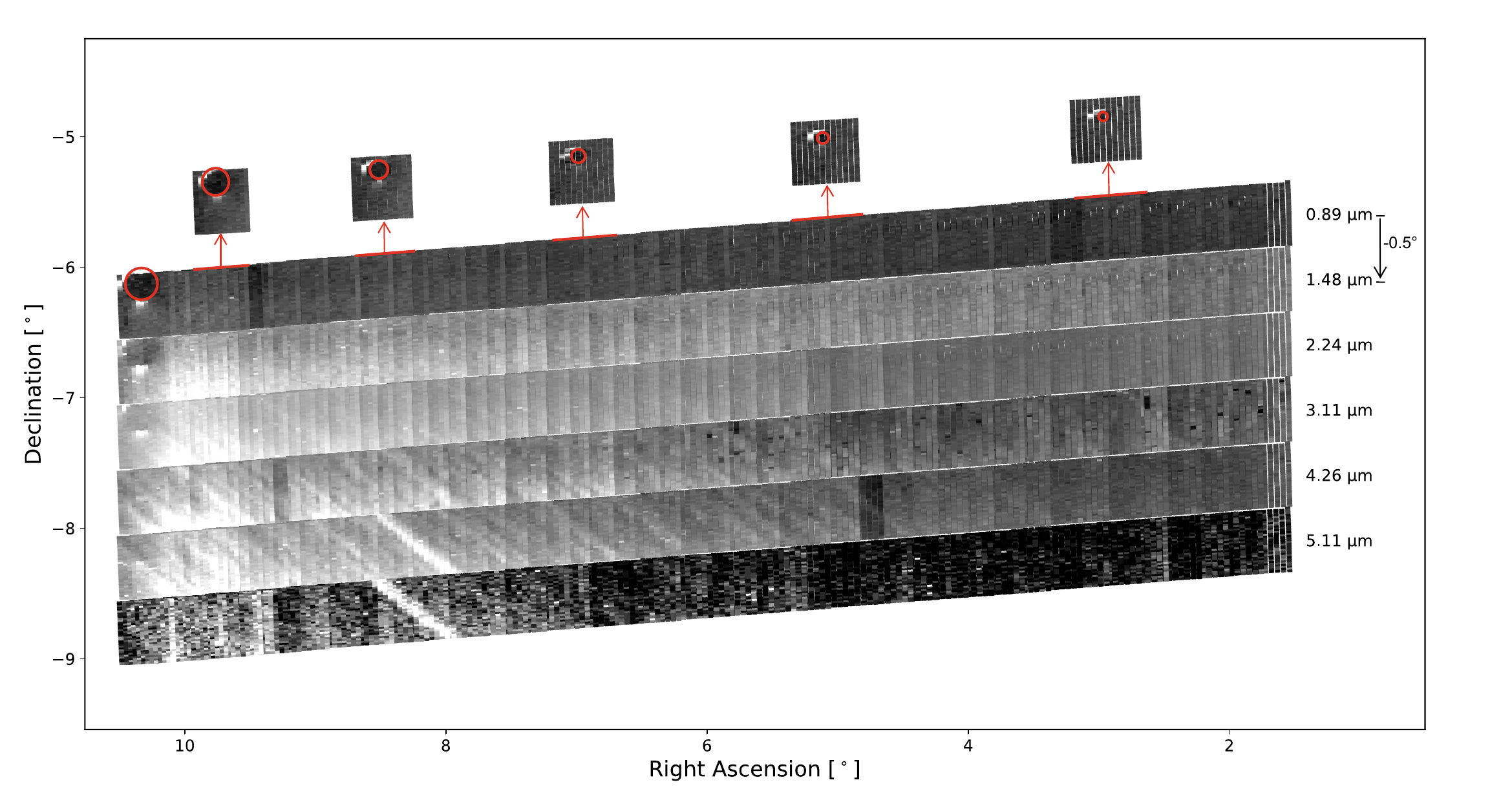}
	\caption{Cassini flyby E17 VIMS mosaic in terms of right ascension and declination in the J2000 reference frame for a discrete set of wavelengths. The original mosaic is given on top; every following wavelength mosaic is shown with an additional declination offset of $-0.5^\circ$ for clarity. The body of Enceladus is marked by a red circle for reference.}
	\label{FIG:4}
\end{figure*}

\noindent Similarly to the ISS observations, VIMS cubes were captured during the entire flyby E17 approach. These are mosaicked according to their $RA$ and $DEC$ coordinates in the J2000 reference frame as shown in \autoref{FIG:4}. This figure shows every cube from the beginning of the E17 flyby ($RA\approx1^\circ$) up to the last cube $\mathrm{1711553290}$ ($RA\approx11^\circ$) for a total of 56 cubes. Observations start at a phase angle of 154$^\circ$ and a distance of $3.4  \times 10^5$ km from Enceladus, which evolve to a phase angle of 160$^\circ$ and a distance of $1.2 \times 10^5$ km at the final cube. During the approach, VIMS continuously tracks Enceladus, ensuring the moon remains in roughly the same pixel location relative to the instrument for each observation. This is illustrated by a few sample cubes extracted from the mosaic, where Enceladus' body is marked with a red circle. In the mosaic, each consecutive image is overlaid onto the previous one. As a result, Enceladus is only visible in the final image, although it is present in each cube near the top centre of the image window.

The flyby mosaic is shown for a discrete set of wavelengths (indicated on the right). The chosen wavelengths are based on milestones in the VIMS-IR spectral range, which represent a significant shift in the cube appearances. The original projection on the sky is shown at the top for $\lambda=0.89$ \textmu m, plotting each subsequent wavelength with an additional declination offset of $\mathrm{-0.5}^\circ$ for comparison. In reality, observations for every wavelength have the exact same projected location in the mosaic. The pixel intensities are normalised over all E17 cubes at a single wavelength, providing a relative view of the intensity throughout the flyby.

Starting from the shortest available wavelength channel of 0.89 \textmu m, no stripes can be seen throughout the flyby. However, the pixel region around $RA=9.8^\circ$ and $DEC=-6.5^\circ$ shows a faint brightness increase over multiple images. This region becomes increasingly bright as the wavelength increases. By the 1.48 \textmu m channel (shown on the second row in \autoref{FIG:4}), this bright area has grown to be approximately as bright as Enceladus' plumes. 

The first stripes appear at $\lambda=2.24$  \textmu m (third row). These parallel stripes become more visible and brighter as Cassini approaches Enceladus, starting from $RA>5^\circ$. The bright area around $RA=9.8^\circ$ remains, exhibiting the most visible stripes. It is important to note that the stripes are continuous over multiple overlapping cubes. As the wavelength increases, the stripes become brighter and more distinct, as shown for $\lambda= 3.11$ \textmu m (row 4). At this point, the bright area is much brighter than the plumes of Enceladus. Additionally, when comparing rows 3 and 4, it becomes evident that the position and brightness of the stripes change with wavelength.

The stripes remain visible at the longer wavelength of $\lambda=4.26$  \textmu m (row 5), with continued variations in position and brightness. A single bright stripe at $RA=8.2^\circ$ stands out, also faintly visible at $\lambda=3.11$ \textmu m. Finally, at $\lambda=5.11$ \textmu m (row 6), the end of VIMS-IR's available spectrum, the data becomes increasingly noisy, and most stripes fade away. However, the single bright stripe at $RA$ = $8.2^\circ$ persists. Regardless of wavelength, all stripes in the mosaic are parallel and inclined to the ecliptic by approximately $16^\circ$ which is the same inclination observed in the ISS mosaic.

\subsubsection{Comparison between ISS and VIMS Stripes}

\begin{figure}
	\centering
		\includegraphics[width=\linewidth]{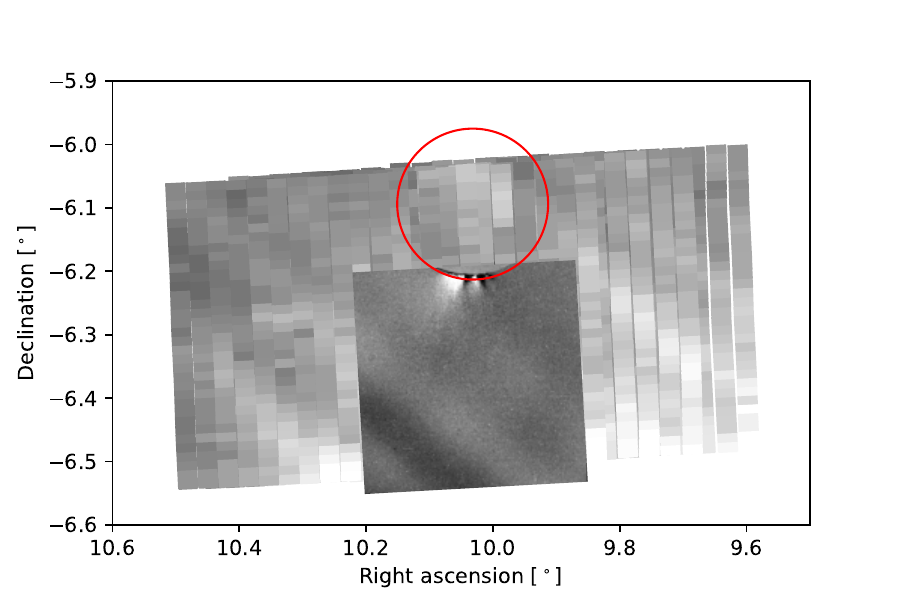}
	\caption{Cassini flyby E17 VIMS and ISS combined mosaic in terms of right ascension and declination in the J2000 reference frame. ISS NAC image N1711552162 taken in IR3 filter and subtracted by closest geometry CLEAR filter image with VIMS cube $\lambda$ = 3.11 \textmu m. The body of Enceladus is marked by a red circle for reference.}
	\label{FIG:5}
\end{figure}

\noindent Unlike the few stripes observed in the ISS data, clear stripes appear in almost every VIMS cube captured during the E17 approach. In \autoref{FIG:5}, a combined mosaic of ISS NAC image N171155216 and four VIMS-IR cubes captured around the same time during flyby E17 is shown. This figure proves that the stripe direction is the same in the VIMS and ISS observations. The stripe width is also similar, with an angular size range of 0.5 to 1.5 mrad. This is comparable to the apparent angular diameter of the Sun at Cassini's position, which is approximately 1 mrad. 

As the VIMS and ISS instruments have completely separate optics \citep{Brown2004TheInvestigation, Porco2004CASSINISATURN}, the stripes cannot be caused by stray light originating from within the instruments. Therefore, it must have an external origin. This conclusion is further supported by the continuous appearance of the stripes across images, a characteristic not typically associated with stray light. In-flight calibration of the ISS-NAC revealed a prominent stray light artefact that appears as a diagonal streak of light (see Figure 18 from \cite{West2010In-flightCameras}). This artefact originates from a bright object just outside the ISS NAC's field of view, like a moon. However, this artefact is dependent on the off-sight object's position. In Flyby E17, the position of Enceladus remains fixed in the image window for each observation. Therefore, if a similar artefact occurs, it would also be fixed in the same position relative to Enceladus in each image window. This is not the case. Unfortunately, these results cannot be compared to known VIMS artefacts as no in-flight stray light analysis was performed for this instrument \citep{Brown2004TheInvestigation}.

\subsubsection{Cassini Observation Geometry}

\begin{figure*}[t] 
	\centering
		\includegraphics[width=\linewidth]{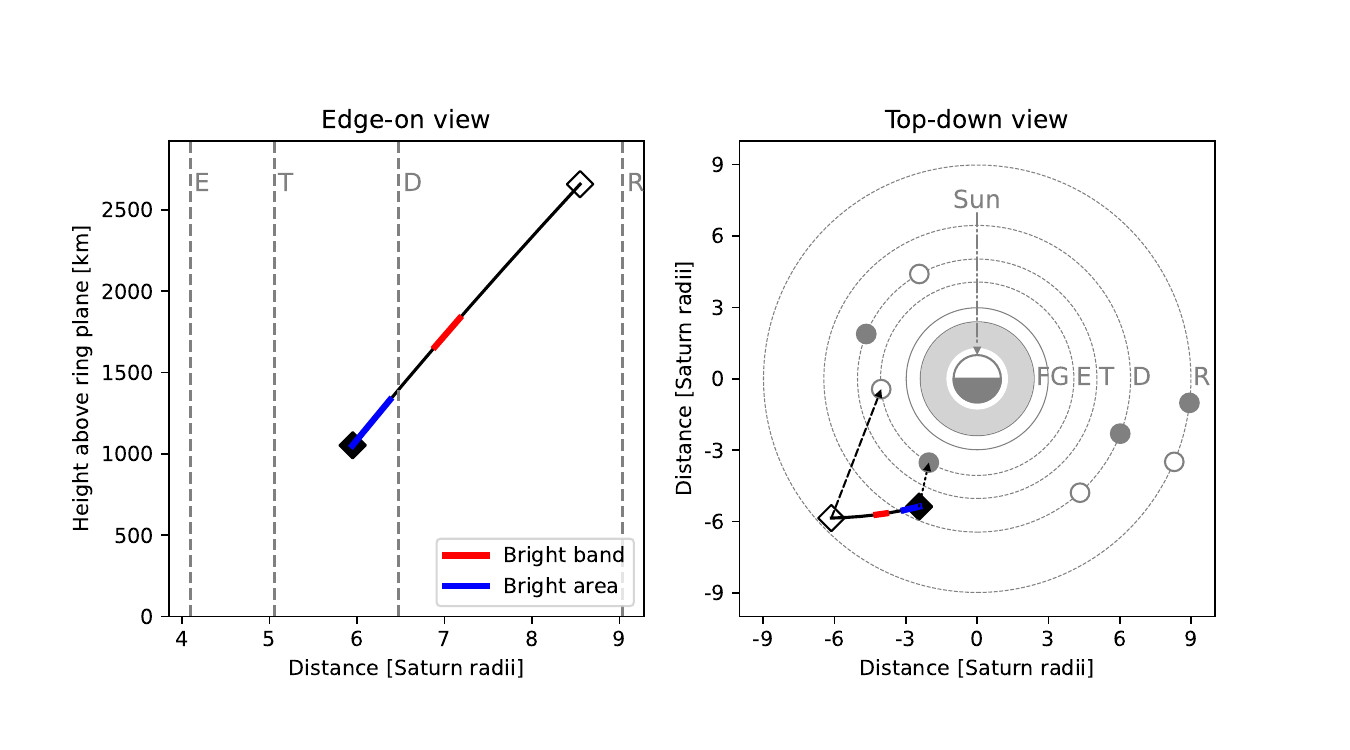}
	\caption{Cassini's E17 flyby geometry on 27/03/2012 from 09:53:14 to 14:49:33 UTC, shown from a Saturn-centred perspective with edge-on (left) and top-down (right) views of Saturn's ring plane. Cassini's position is marked by a diamond, and the moons' positions are indicated by circles. The open symbol represents the start position, and the filled symbol represents the end position. The orbits of Enceladus, Tethys, Dione, and Rhea (E, T, D, R) are represented by dashed grey lines, while the F ring and G ring (F, G) are shown with solid grey lines. The solar direction is indicated by a dash-dotted arrow, and the ISS/VIMS viewing direction by black dashed arrows.} 	\label{FIG:2d_geo_E17}
\end{figure*}

\noindent An overview of the observation geometry of Cassini during flyby E17 is shown in \autoref{FIG:2d_geo_E17}. Starting from the edge-on view of Saturn's ring plane, it can be seen that at the beginning of the approach, Cassini is at a distance of 8.5 $\mathrm{R_S}$ (Saturn Radii) and at an altitude of 2600 km above the ring plane. During its approach, Cassini moves closer to Saturn and the ring plane. The observation window of the bright band observed by six VIMS cubes in \autoref{FIG:4} is shown in red. The bright band appears for a range of $6.9<\mathrm{R_S}<7.2$, before crossing Dione's orbit. Shortly after this crossing, the bright area appears and remains visible till the end of the imaging approach. A top-down view is shown on the right side of \autoref{FIG:2d_geo_E17}, revealing the observed sections of Saturn's inner moon system. Note that the Sun's elevation above the ring plane is not zero but 13.7$^\circ$ North from the centre of the ring plane. The bright stripe is visible when the F ring should appear in the background of the field of view. In addition, the G ring and other inner rings (indicated by the grey area) can also be present in the background of images after the bright band observations.

\subsection{Search Methodology for Similar Stripe Observations}
\label{subsec:search_method}
\noindent To investigate the nature of the stripes it is important to consider their possible appearance in other observations. To this end, Cassini ISS NAC and VIMS-IR observations with similar observation and illumination conditions to flyby E17 were checked. The image search is separated into two components: images containing Enceladus' plumes and images of other targets (without plumes).  

The selection criteria for the ISS NAC images flybys are based on the phase angles of the E17 flyby images where stripes were observed. Specifically, the stripes appeared at phase angles ranging from 160.2$^\circ$ to 160.4$^\circ$. Consequently, flybys with phase angles within this range with an added boundary $\pm 2^\circ$ are included in this analysis. This boundary results in conservative geometry requirements while increasing the number of observations (previously 80). The ISS NAC images in each flyby are first analysed by applying a linear stretch of the $I/F$ values so that the plumes are saturated. This method reveals the faint stripes in flyby E17 as shown in \autoref{FIG:1}. Secondly, observations in the CLEAR filter are subtracted from images captured with medium/broad-band filters (if available). This method reveals clear stripe signatures without manually adjusting the pixel intensity scale. No stripes in ISS NAC images were found after applying both checks. 

Stripes in the VIMS cubes can be identified more straightforwardly by visually inspecting cube previews available in the VIMS Data Portal from \cite{LeMouelic2019TheMaps}. The stripes observed in \autoref{FIG:4} from flyby E17 are clearly visible in these previews, which show the cubes at wavelengths of 1.78, 2, 3.1, 3.45, and 5 \textmu m. As a result, this method proves to be a quick and efficient way to locate similar stripes across other targeted flybys. Consequently, every targeted flyby of Titan, Enceladus, Dione, Rhea, Phoebe, Tethys, Hyperion, and Iapetus was checked for stripes. This analysis identified similar stripe features in VIMS-IR from two additional Enceladus-targeted flybys: E13 and E19. Further details along with the projection mosaics and geometry analysis for these flybys can be found in the supplementary material. The stripes observed in flybys E13 and E19 exhibit similar widths (0.6 to 2.6 mrad) and chromatic characteristics to those in E17. Additionally, all stripes are inclined at $16^\circ$ to the ecliptic and remain continuous across multiple cubes. These consistent properties across flybys E13, E17, and E19 strongly suggest a common origin for the observed stripes.

Candidate stripes with differing appearances were also found. Examples include cubes targeting Dione, Rhea, Titan, and Tethys, as shown in Figure 4 from the supplementary material. These show bright parallel stripes at similar wavelengths to the stripes in E17. However, either not enough observations are available to mosaic the stripes or the stripes do not align between images. Additional key differences in appearance are stripes that are strictly localised to the lowest-phase corner of the image, an additional set of stripes perpendicular to the other stripes, and stripes that are perfectly parallel to the sun-direction vector (flyby E17, E13 and E19 show an off-set to this vector). Therefore, these observations are disregarded in the subsequent analysis.

\subsection{Stripe Observation Summary}
\label{subsec:summary_obs}

\noindent An overview of the stripe observations is provided in \autoref{tbl2}, encompassing four independent detections by the Cassini ISS and VIMS instruments. The recorded stripes are observed during high-phase, Enceladus-targeted observations, all showing a consistent inclination of $16^\circ$ relative to the ecliptic. The simultaneous detection of the stripes in both ISS and VIMS during Flyby E17, along with the continuity of the stripes between the mosaicked images, suggests an external origin rather than stray light. The stripes' angular widths vary from 0.5 to 2.6 mrad. These may appear brighter and in front of Enceladus. Their visibility depends on the observed wavelength, with peak visibility in the VIMS data at wavelengths between 2.2 and 4.3 \textmu m for all flybys. Notably, stripes detected in VIMS-IR do not necessarily correspond to stripes in the ISS NAC under the same observation geometry. This may be due to reduced stripe visibility at wavelengths below 2.2 \textmu m (as seen in VIMS flybys E13, E17, and E19) combined with a phase angle dependence, where stripes are visible only at phase angles greater than 160$^\circ$ (see supplementary material). A distinct bright stripe, referred to as the "bright band", emerges in flyby E17 at $\lambda >$ 4.26 \textmu m, which stands out from the other stripes. The observations, spanning up to 1.37 years apart, suggest that the stripe phenomena may persist for at least that duration.

\begin{table}[width=\linewidth,cols=4,pos=h]
\caption{Summary of stripe observations.}\label{tbl2}
\begin{tabular*}{\tblwidth}{@{}L L p{3cm} L@{} } 
\toprule
Flyby & Instrument & Time Range [UTC] & Phase Range [$^\circ$]\\
\midrule
E13 & VIMS-IR & 2010-12-20 22:53:39--2010-12-21 00:00:40 & 164.0--165.0\\
E17 & VIMS-IR & 2012-03-27 09:53:14--14:43:56 & 154.0--160.3 \\
E17	& ISS & 2012-03-27 14:15:56--14:43:00 & 160.2--160.4 \\
E19 & VIMS-IR & 2012-05-02 05:07:13--06:02:00
 & 159.0--160.0 \\
\bottomrule
\end{tabular*}
\end{table}

\begin{figure}[h]
    \centering
    \includegraphics[width=\linewidth]{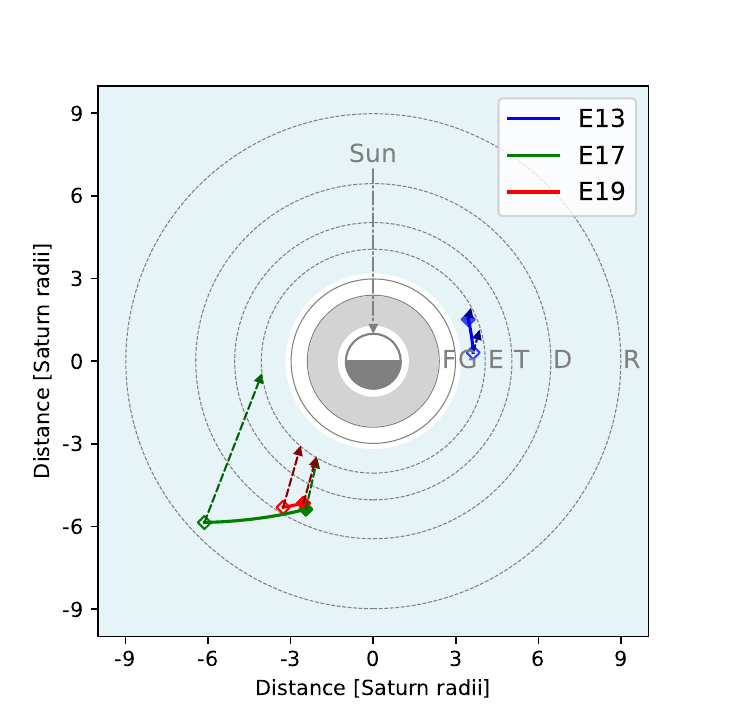}
    \caption{Observation geometry summary of flybys containing VIMS stripe features represented in a top-down view of Saturn's ring plane. The extent of the E ring is shown in light blue.}
    \label{fig:2d_geo_sum}
\end{figure}

A summary of the top-down geometry of the VIMS stripe observations relative to the Saturn system and aligned with Saturn's solar illumination is presented in \autoref{fig:2d_geo_sum}. All observations were conducted when Cassini was near the ring plane and within the extent of the E ring (from Mimas' orbit to Titan's \citep{Horanyi2008Large-scaleE-ring}). Flybys E17 and E19 exhibit very similar viewing geometries, with Cassini positioned between the orbits of Tethys and Rhea, facing Saturn. In contrast, flyby E13 features a different geometry, with Cassini located between the G ring and Enceladus' orbit, this time looking away from Saturn.

From the flyby mosaic shown in \autoref{FIG:4}, it became apparent that the stripes are fixed in RA and DEC space. Yet, Enceladus's location is not; it traces its orbit with time, as seen from the Enceladus-centred VIMS cubes (each cube contains Enceladus in the same area of the viewing window). Therefore, the observed stripe phenomenon occurs in a system decoupled from the motion of Enceladus and can thus not occur due to a physical structure which co-rotates with Enceladus orbit. This means that the origin of the stripes likely lies in front or behind Enceladus along Cassini's line of sight.

\section{Stripe Characterisation}
\label{sec:character}

\noindent The objective of this section is to characterise the stripe features observed in the Cassini ISS and VIMS data. This will involve examining the spectral variations in the appearance of the stripes, leading to the identification of the source of the stripe features. Subsequently, we will analyse the material responsible for the formation of these stripes.

\subsection{Spectral Variations}

\noindent The ISS images and VIMS cubes show spectral dependencies in the intensity and positioning of the stripes. This section aims to analyse these effects and identify their origin. We will start by examining spectral variations in individual observations, followed by a comparative analysis across multiple flyby images.

\subsubsection{Stripe Spectral Dependencies within Single ISS and VIMS Images}

\noindent We compare the spectral variations between the simultaneous flyby E17 ISS NAC and VIMS stripe observations. For the ISS these include the last sequence of images corresponding to the last four entries in \autoref{tbl1} and the left-most images in each filter in \autoref{FIG:2}. The VIMS cube 1711553290\_1 previously shown in \autoref{FIG:3} is used for comparison with a mid-time between the IR3 and IR1 ISS images. 

\begin{figure}[h]
	\centering
		\includegraphics[width=\linewidth]{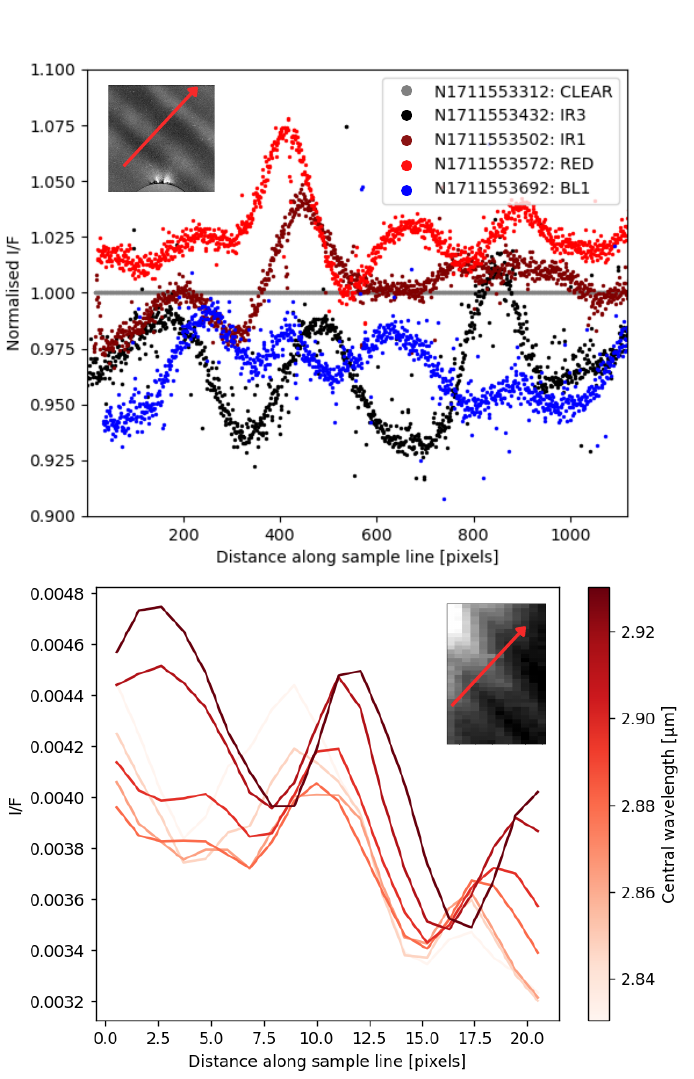}
	\caption{Intensity curves of ISS NAC sequence 12 (top) and VIMS cube 1711553290 (bottom) along a sample line indicated by the red arrows. The ISS $I/F$ values are normalised by CLEAR filter images. The VIMS $I/F$ samples are shown for spectral bands with varying central wavelengths and a bandwidth of 50 nm. }
	\label{FIG:6}
\end{figure}

The intensities of the respective images along a line perpendicular to the direction of the stripes are sampled. This sampling line is marked with a red arrow in \autoref{FIG:6}. The intensity curves from the ISS, displayed at the top of the figure, are adjusted for pixel drift because each ISS image was taken at a different time and with a different filter. This introduced slight shifts between the image projections, as seen in the mosaic in \autoref{FIG:2}.

The ISS NAC intensity samples reported in \autoref{FIG:6} show a bright peak around the 400 to 500-pixel mark for every filter. This is the bright diagonal stripe feature from \autoref{FIG:1}. It can be seen that these peaks are slightly offset from each other. In addition, there is a continuous shift to the right for these intensity peaks with increasing wavelength. Therefore, an increase in $\lambda$ results in the bright diagonal stripe moving to the top-right along the direction of the sample line.

The VIMS intensity curves for different wavelengths are examined at the bottom of \autoref{FIG:6}. The line samples are obtained using linear interpolation of the pixel values along the line and then processed through a moving average smoothing algorithm to achieve smooth intensity curves despite the limited number of available pixels. Noise in the data is further reduced by sampling the average intensity of spectral bands with a bandwidth of 50 nm (4 VIMS channels) instead of a single channel. Starting from the shortest wavelength at 2.81 \textmu m in pale pink, a peak at a sampling line distance of 9 pixels appears. With increasing wavelength, this peak shifts further to the right along the sample line and decreases in intensity until 2.88 \textmu m (shown in orange). Increasing the wavelength further, the intensity increases again, and the peak continues to shift towards the right. Extending the analysis to the neighbouring VIMS channels results in the same periodic intensity fluctuations and the continuous shift in stripe spatial position with increasing wavelengths. Additionally, the periodic intensity fluctuations follow a sinusoidal pattern, where the global intensity decreases with increasing wavelength while the amplitude remains constant. The direction of this spatial shift is the same for the ISS and VIMS images. 

To conclude, it is shown that the stripes features seen in both VIMS and ISS experience a continuous shift in positions perpendicular to their tilt when observed at different wavelengths. This chromatic shift is paired with a periodic stripe intensity fluctuation following a sinusoidal pattern.

\subsubsection{Retrieval of Spectral Structures Throughout The Flyby}

\begin{figure*}[T]
    \centering
    \includegraphics[width=\linewidth]{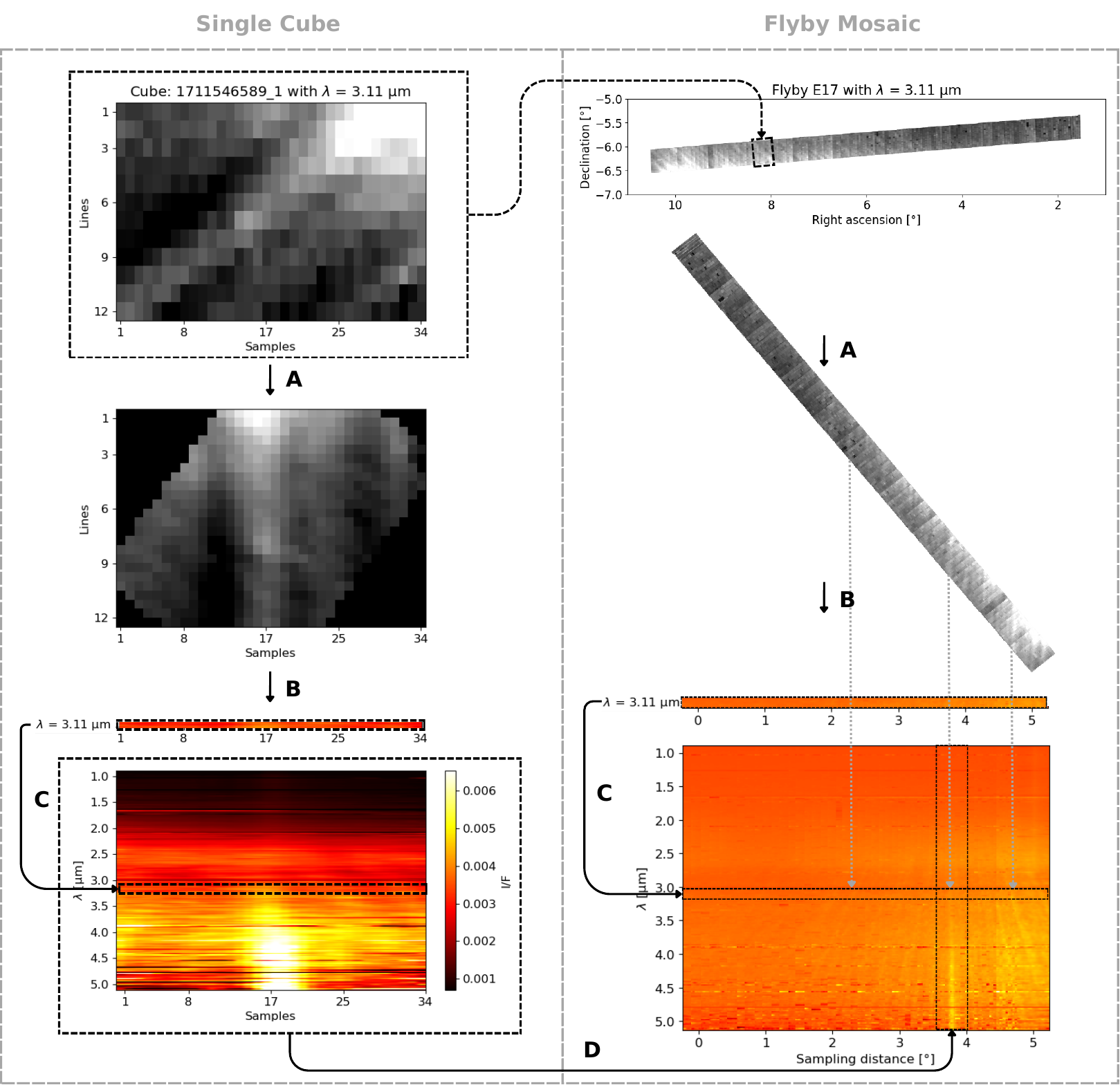}
    \caption{Flow diagram for the creation of a Cassini VIMS flyby spectral kymograph.}
    \label{FIG:kymo_method}
\end{figure*}

\noindent After observing the localised spectral shifts of the stripes, we aim to map these variations across the entire flyby. To achieve such a map, spectral information is extracted from VIMS cubes using the methodology outlined in \autoref{FIG:kymo_method}. This method adapts a graphical representation from cell biology known as a kymograph \citep{Nitzsche2010ChapterAssays}. Kymographs are space-time plots that depict intensity values along a predefined path over time by trading one spatial dimension for a temporal one. However, this analysis replaces the temporal axis with a spectral dimension. 

The process begins by considering a single VIMS cube observed in a specific wavelength channel, as depicted in the top left of \autoref{FIG:kymo_method}. In step \textbf{A}, the image is rotated so that the VIMS cube stripes align vertically with the observation window. The tilt of the stripes with respect to the image horizontal (samples axis) differs for each flyby. Yet, because of Cassini's varying orientation, they all result in a 16$^\circ$ tilt to the ecliptic when mosaicked in the J2000 reference frame. The rotation angles for vertical stripe alignment in the cubes are 49.8$^\circ$, 54.2$^\circ$, and 47.8$^\circ$ for flybys E13, E17, and E19, respectively. In step \textbf{B}, the pixels within a single column of the rotated image are averaged, resulting in a single line where the horizontal axis represents the physical dimension of the cube’s field of view. The vertical axis now corresponds to a single wavelength. This procedure is repeated across the 256 VIMS-IR channels for the same cube. The resulting lines are then stacked to encompass the full VIMS-IR wavelength range on the vertical axis, as shown in step \textbf{C}. The result is a spectral kymograph for a single cube, with the horizontal axis representing the number of pixels (called samples) and the vertical axis representing wavelength. In this kymograph, continuous structures display the gradual displacement of the stripes across different wavelengths. For example, the bright stripe at sample 17, observed in the cube at $\lambda=3.11$ \textmu m, intensifies as the wavelength increases, while its position remains unchanged along the sample direction. Another stripe can be seen from a high-intensity region starting from $\lambda=3.11$ \textmu m at sample 24. The bright region gradually shifts to the right with increasing wavelength along the sample direction, indicating a rightward displacement of the stripe.

This methodology can also be extended to an entire flyby mosaic, as illustrated starting from the top right of \autoref{FIG:kymo_method}. Initially, the mosaic is rotated, and then the column averages are retrieved for each wavelength channel. These averages are then stacked to generate a spectral kymograph mosaic. However, the horizontal physical dimension now depicts an angular distance projected onto the celestial sphere. Step \textbf{D} demonstrates how an individual image cube is integrated within the kymograph mosaic. In practice, the kymograph mosaic is constructed by superimposing each individual cube kymograph onto the next in a manner analogous to the creation of the cube mosaics. The intensity values in each cube kymograph are normalised over the entire flyby, such that an $I/F$ value of 0 corresponds to the global minimum, and an $I/F$ value of 1 represents the global maximum throughout the flyby.

The bottom right plot in \autoref{FIG:kymo_method} presents a spectral kymograph mosaic after applying the methodology above on flyby E17. A single vertical bright line can be seen at a sampling distance of 3.85$^\circ$. This line becomes visible at $\lambda=3$ \textmu m and gains contrast as $\lambda$ increases, standing out as the sole bright feature at $\lambda=5$ \textmu m. The vertical alignment indicates that the feature's position remains constant across all wavelengths. These characteristics correspond to the bright band described in \autoref{FIG:4}. The bright area can also be observed in the kymograph with a global brightness increase after a sampling distance of 4.4$^\circ$. Other notable features are the multiple slanted lines symmetrically arranged around the bright band. These lines suggest a physical phenomenon where the stripes observed in \autoref{FIG:4} shift symmetrically and continuously away from the bright band as $\lambda$ increases.

The spectral shift of the stripes observed in flyby E17 resembles the diffraction of light. Assuming that there is an ideal diffraction grating with plane light waves propagating normal to the grating, the grating intensity $I$ as a function of the diffracting angle $\theta_d$ is given by \autoref{eq:grating} \citep{Thorne1988DiffractionGratings}. The first term represents the Fraunhofer diffraction distribution for a single slit with width $b$. The second term equals the contributions from $N$ slits with equal spacing $d$.

\begin{equation}
I(\theta_d) = \left( \frac{\sin (\beta)}{\beta} \right)^2 \left( \frac{\sin (N \alpha)}{\sin (\alpha)} \right)^2 \\
\text{where} \quad
\begin{aligned}
\beta &= \frac{\pi b \sin (\theta_d)}{\lambda} \\
\alpha &= \frac{\pi d \sin (\theta_d)}{\lambda}
\end{aligned}
\label{eq:grating}
\end{equation}

\begin{figure}[h]
    \centering
\includegraphics[width=\linewidth]{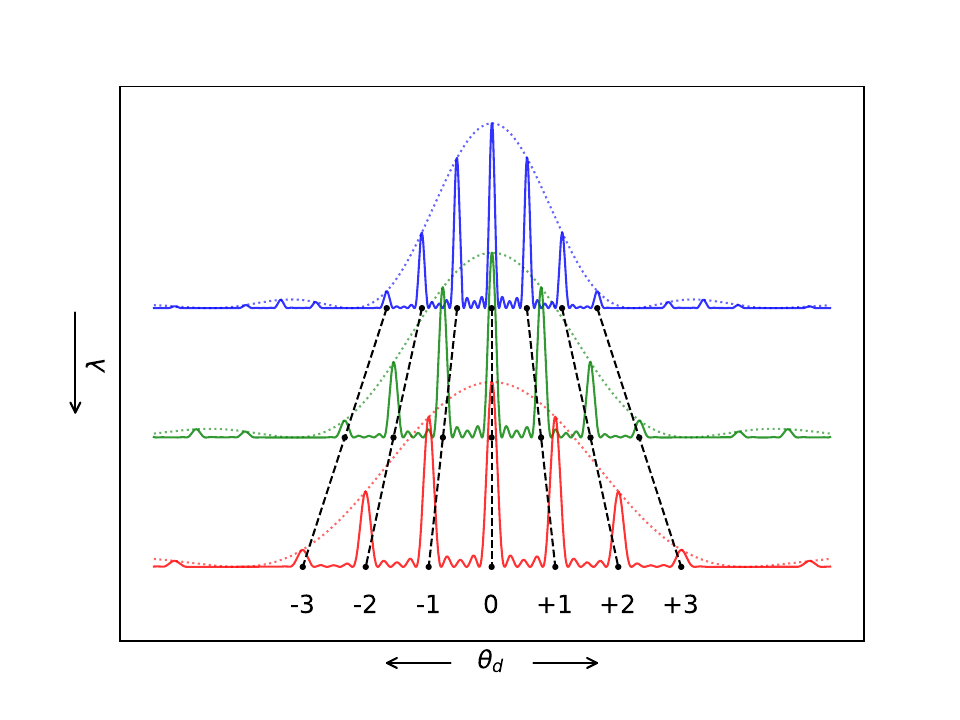}
    \caption{Diffraction pattern from a finite grating with 5 slits and $b=d/4$ in relation to diffraction angle and wavelength similar to the spectral kymograph.}
    \label{fig:diff_pattern}
\end{figure}

\autoref{fig:diff_pattern} illustrates the resulting intensity patterns for a grating with five slits for three discrete wavelengths. The diffraction angle is represented on the x-axis, while the wavelength is plotted on the y-axis, similar to the spectral kymographs. The dotted curves show the intensity envelope for a single slit, and the full line curves include the contributions of the five slits. The result is a symmetric diffraction pattern for a given $\lambda$, with a $0^{\mathrm{th}}$-order diffraction peak at $\theta_d=0$ and higher order peaks emerging from constructive interference at larger $|\theta_d|$. \autoref{fig:diff_pattern}, represents a simplified case. In reality, the $0^{\mathrm{th}}$-order peak cannot be created by forward diffraction from a transmission grating but must be produced by reflection because Cassini observes the bright band at a phase angle of $159^\circ$ rather than $180^\circ$. Therefore, a reflection grating consisting of an orderly-spaced reflecting surface must create the diffraction pattern.

\cite{Thorne1988DiffractionGratings} proved that the $\theta_d$ for a diffraction peak or maxima of integer order $m$ created from a grating with slit spacing $d$ at a particular $\lambda$ is determined according to the diffraction grating equation shown in \autoref{eq:diffract}. $\theta_i$ is the incidence angle of the plane wave with respect to the grating normal. For the $0^{\mathrm{th}}$-order, $\theta_i = -\theta_d$, making it independent of $\lambda$. This corresponds to specular reflection in a reflection grating.

\begin{equation}
\sin \theta_i+\sin \theta_d=\frac{m \lambda}{d}
\label{eq:diffract}
\end{equation}

\autoref{eq:diffract}, can be adapted to get $\sin \theta_d-\sin \theta_0=\frac{m \lambda}{d}$ which is the angle between the maxima of the $m$ order and the $0^{\mathrm{th}}$-order. This linear relation between $\theta_d$ and $\lambda$ for a discrete set of $m$ is presented in \autoref{fig:diff_pattern} by the black dashed lines with the orders annotated. These connect the same order peaks for diffraction patterns created by multiple wavelengths but with the same $d$. The result is a pattern of slanted lines similar to those observed in the spectral kymograph.

In the case of a single slit, represented by the dotted curves in \autoref{fig:diff_pattern}, there is a substantial fall intensity for higher-order maxima which conforms to the diffraction hypothesis as the slanted lines in the kymograph are less intense than the bright band. However, orders beyond the third contribute only marginally to the intensity. In contrast, the kymograph shows several distinct lines of roughly equal intensity at various scattering angles. The presence of multiple slits could explain this behaviour. A conservative range of slit widths can be obtained from \autoref{eq:diffract} by assuming that the brightest of both the closest and furthest lines from the bright band are created from a 1$^{\mathrm{st}}$-order diffraction peak. The nearest line is observed at an angular distance of 0.113$^\circ$ from the bright band at $\lambda = $ 5.11 \textmu m, resulting in an upper groove spacing limit of 2.60 mm. Identifying the most distant and brightest line from the bright band is challenging due to the similar brightness levels of the lines. In this instance, the stripe crossing the $\lambda = $ 3 \textmu m dip at a sampling distance of 2.4° is selected as the most distinguishable. This gives a lower limit for the groove spacing of 0.12 mm. Consequently, the expected groove spacing range for the reflection grating is 0.12–2.6 mm.

Spectral kymographs are also created for the E13 and E19 VIMS mosaics. These, along with a detailed analysis, are included in the supplementary material. No vertical or slanted lines are observed in flyby E13. However, E19 exhibits distinct slanted lines, which behave as expected when comparing the flyby geometries and kymographs between E17 and E19.

\subsection{Characterisation of the Material at the Origin of the Stripes}
\label{sec:ices}

\noindent To investigate the nature of the particles that create the bright band, VIMS-IR spectra can be retrieved and compared to other features like the plumes and the E ring background. Noise is prominent for high-phase observations of faint structures like the plumes \citep{Hedman2009SpectralCassini-vims}. Therefore, spectra of five cubes containing the bright band were averaged to extract meaningful spectra. Details regarding these observations are shown in \autoref{tbl3}.   

\begin{table}[width=\linewidth,cols=4,pos=h]
\caption{Cubes containing the bright band.}\label{tbl3}
\begin{tabular*}{\tblwidth}{@{}L L L L@{} } 
\toprule
Cube ID & Cube Mid-Time [UTC]& Range [km] & Phase [$^\circ$]\\
\midrule
1711545954\_1 & 2012-03-27 12:36:53  & 201,832  & 158.6\\
1711546301\_1 & 2012-03-27 12:42:40 & 197,424  & 158.8\\
1711546589\_1 & 2012-03-27 12:47:28 & 193,813 & 158.9\\
1711546939\_1 & 2012-03-27 12:53:18 & 189,440  &159.0\\
1711547227\_1& 2012-03-27 12:58:06 & 185,890 & 159.1\\
\bottomrule
\end{tabular*}
\end{table}

\begin{figure*}[T]
    \centering
    \includegraphics[width=\linewidth]{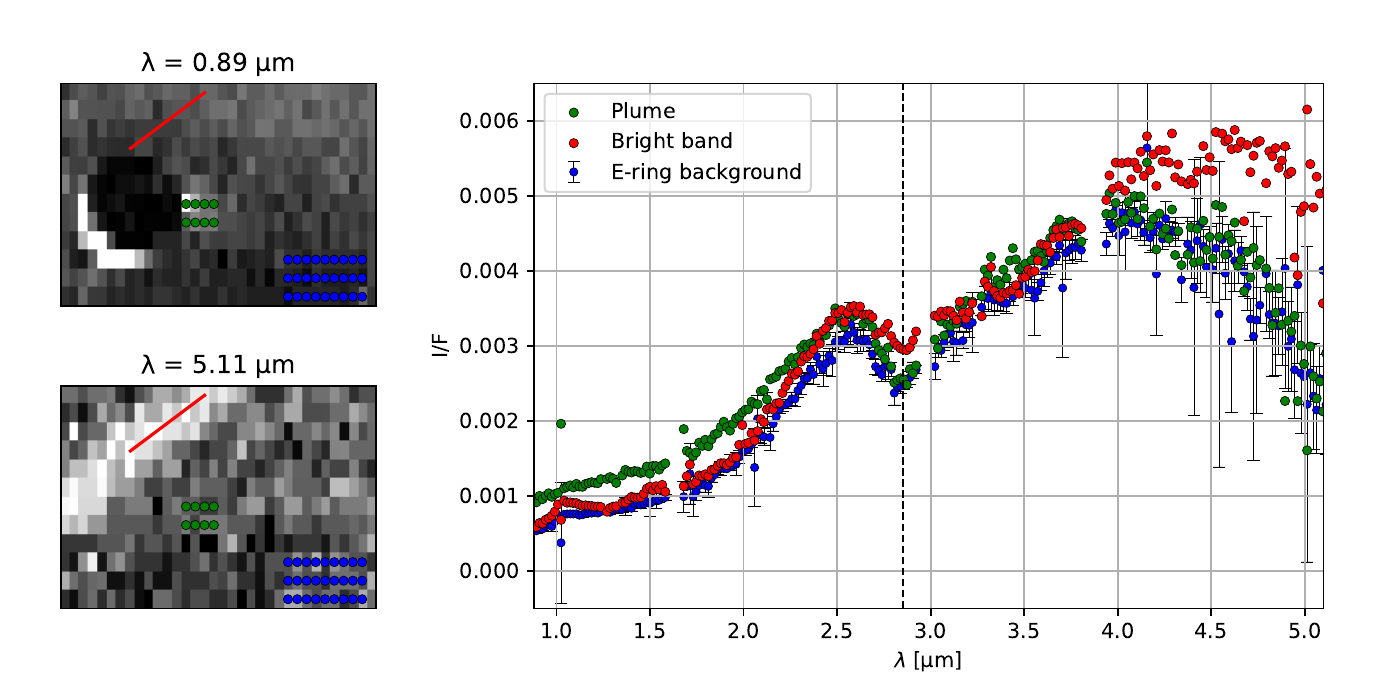}
    \caption{(left) Example of sampling areas of VIMS cube 1711547227\_1, (right) spectra averaged over all bright band observations. The dashed line indicates the position of 2.85 \textmu m for comparison.}
    \label{fig:spec}
\end{figure*}

An example of the spectra retrieval inspired by the analysis of \cite{Dhingra2017SpatiallyVIMS} is illustrated in the left part of \autoref{fig:spec}. The bright band spectrum is extracted by averaging pixels along a line (shown in red) using spline interpolation on the pixels values in the line's vicinity. The bottom figure shows that the line does not extend the entire length of the bright band. This is because pixels in the bright band that correspond to the plumes or body of Enceladus are ignored. This is to prevent other contributions to the bright band spectra. Next, the spectra of the plumes are extracted by averaging the pixel intensities over the area shown in green. Similarly, the E ring background is determined by averaging the pixel values in blue. The process is repeated for each of the five cubes containing the bright band. Note that in each image, the bright band is in a different location, therefore, the line samples are adjusted for each image. Sample locations for the plumes and the E ring background remain the same, as justified by the similar observation geometries. A measure for the noise in the cubes is obtained by computing the standard deviation of the pixels used to calculate the E ring background measurements similar to the method of \cite{Dhingra2017SpatiallyVIMS}. Additionally, data from known hot channels (1.24 \textmu m, 1.33 \textmu m, 3.23 \textmu m, 3.24 \textmu m, 3.83 \textmu m) and from IR focal plane blocking filters (1.6–1.67 \textmu m, 2.94–3.01 \textmu m, and 3.83–3.89 \textmu m) is excluded in this analysis \citep{Brown2004TheInvestigation, Nicholson2008AVIMS}. The wavelength uncertainty for VIMS-IR channels is less than 10 nm \citep{Brown2004TheInvestigation}.

Distinct differences between the plume, the bright band, and the E ring background appear from the averaged spectra on the right of \autoref{fig:spec}. These are analysed through indicators inspired by the analysis of \cite{Dhingra2016NearControls}.

\textit{(A) 1--2 \textmu m \&  2--2.6 \textmu m Spectral Slopes:} The bright band spectrum is similar to the E ring background for $\lambda<$ 2 \textmu m aside from two smooth peaks centred around 1 and 1.6 microns. The plume is generally brighter in this spectral range with a similar spectral slope as the bright band and the E ring. In the spectral range of 2--2.6 \textmu m, the bright band spectrum has a steeper slope than the plume and E ring spectra. Because the wavelength range of the observations is related to the particle sizes in the scene, the spectrum of the bright band suggests a particle size distribution with a higher percentage of larger particles compared to the plumes and E ring background \citep{Hedman2009SpectralCassini-vims, Dhingra2016NearControls}.

\textit{(B) 3 \textmu m Water-Ice Band Position}: The position of the band minimum of the 3-micron water absorption band is identical for the plumes and the bright band, which is just to the right of the 2.85 \textmu m mark shown by the dashed vertical line. The E ring background spectrum has a minimum below 2.85 \textmu m. \cite{Mastrapa2008Opticalm} showed experimentally that the position of the band minimum can be used to distinguish amorphous and crystalline water ice. As the ice transitions from amorphous to crystalline, the band minimum shifts approximately 50 nm towards longer wavelengths. Therefore, the plume consists of mainly crystalline ice while the E ring contains more amorphous ice, which agrees with the results from \cite{Dhingra2017SpatiallyVIMS}. The bright band has the same ice crystallinity as the plumes.

\textit{(C) 3 \textmu m Water-Ice Band Shape}: The shape of the 3-micron water absorption band is similar for the plume and the E ring background. However, the absorption feature is less steep and narrow for the bright band, with an added bump close to 3 \textmu m. \cite{Dhingra2017SpatiallyVIMS} found that the 3-micron water absorption band shape strongly depends on particle size distributions. This bright band spectrum could be the result of a paucity of particles with a radius smaller than 2 \textmu m. The plume spectrum shows the greatest band depth, followed by the E ring and the bright band. This indicates that water ice is most abundant in the plume and least abundant in the bright band's material.  

\textit{(D) > 4 \textmu m Spectral Slope \& CO$_2$ Absorption:} The signal-to-noise ratio is low in the spectral region of $\lambda > 4$ \textmu m as shown by the large error bars, yet some features can still be discerned. The bright band spectrum is significantly brighter than the plume and E ring background, which was also observed directly from the cubes in this spectral range. This again suggests that larger particles are present in the bright band's material compared to the material in the plumes and E ring. In addition, there is a dip in the bright band spectrum at around 4.27--4.42 \textmu m, which can be associated with the absorption of solid phase CO$_2$ \citep{Postberg2018PlumeEnceladus}. This absorption band is not seen in the other spectra.

In conclusion, the spectrum of the bright band suggests the presence of water ice that is as crystalline as the particles in Enceladus' plumes, though it contains less water than the plumes and E ring. The particle size distribution in the bright band likely includes a higher proportion of larger particles than the plumes and E ring, with fewer grains smaller than 2 \textmu m. Fresh E ring particles supplied by the plumes are generally larger than the older particles in the E ring due to space-weathering effects, such as sputtering \citep{Nolle2023RadialEffects}. Also, space-weathering leads to a decrease in more volatile compounds in the ice, such as CO$_2$ and can transform crystalline ice to amorphous ice \citep{Dartois2015HeavyYield}. As a result, the presence of large crystalline ice particles and solid CO$_2$ in the bright band material suggests that the bright band may consist of very fresh ice which is less affected by space-weathering effects.

\section{Interpretation \& Discussion}
\label{sec:hypo}

\noindent We summarise the observation of stripe observations in flybys E13, E17, and E19. Flyby E17 revealed stripes in concurrent VIMS-IR and ISS measurements, while the other flybys only included VIMS-IR observations. These high-phase observations show continuous stripes between images when projected on the sky in the J2000 reference frame using RA-DEC coordinates. The stripes are all parallel and have an apparent inclination of $\sim$16$^\circ$ to the ecliptic and $43^\circ$ to Saturn's ring plane. The stripes may also appear brighter than Enceladus and are positioned in front of the moon. Interestingly, under the same observation geometry, stripes found in VIMS-IR do not always correspond to stripes in the ISS NAC. A combination of phase angle dependence, where stripes are visible only at phase angles greater than 160$^\circ$, and decreased stripe visibility at wavelengths below 2.2 \textmu m (as observed in VIMS flybys E13, E17, and E19) could be the cause of this.
 
A single bright line observed by VIMS-IR during flyby E17 lies at the centre of these stripe observations. This bright band becomes visible at $\lambda > 3$ \textmu m and increases in contrast up until $\lambda =$ 5.11 \textmu m. The width of this bright band is approximately 1 to 1.5 mrad, which is similar to the apparent width of the Sun as observed by Cassini. The 43$^\circ$ inclination of this feature to Saturn's ring plane is difficult to reconcile with known physical structures in the E ring. \citet{Kempf2010HowRing, Hedman2011TheRing} found that E ring particles launched from Enceladus’ plumes can only reach orbital inclinations up to 1$^\circ$. Impacts of E ring particles on the inner mid-sized moons also contribute material to the E ring \citep{Dobrovolskis2010ExchangeOrbits}, but simulations show similarly low inclinations \citep{Kempf2008TheParticles}. Other material, such as that from Saturn's inner rings, is also unlikely to produce high inclinations since it remains confined to Saturn's ring plane. The close alignment of the stripes with the ecliptic could hint towards a source beyond Saturn's Hill sphere, like interplanetary dust or cometary debris. However, such contributions are relatively small compared to material generated within Saturn’s system \citep{Hsu2011StreamSaturn}. 

Ice crystals may become preferentially oriented with the electric field generated by Saturn's magnetosphere \citep{Saunders1999TheThunderstorms, Jones2006EnceladusSaturn}, potentially creating the necessary inclination. This alignment can also lead to the formation of unusual halo displays \citep{Moilanen2022LightHalos}. However, the intricate interactions between the plumes of Enceladus and Saturn's magnetosphere, along with the uncertainty surrounding the shapes of the plume ice crystals, complicate predictions regarding halo phenomena and crystal orientations. In addition, these ice particle halos require ice crystals that are significantly larger than the observed wavelength \citep{Bi2014AccurateMethod}. However, most particles in the E ring are comparable in size to the observed wavelengths of the ISS and VIMS \citep{Hedman2009SpectralCassini-vims, Ingersoll2011TotalImages}, making diffraction the primary mechanism for light scattering in this context. This aligns with the diffraction behaviour observed in \autoref{FIG:kymo_method}. We found that the stripes display chromatic behaviour similar to a reflection grating where the bright band itself is observed as reflected light, while the other stripes in the flyby are likely created from higher-order diffraction peaks. The estimated groove spacing of the grating should be 0.12 to 2.60 mm.

The spectrum of the bright band's material suggests that it consists primarily of crystalline water ice. The particle size distribution likely contains larger particles than the plume and E ring background or a lack of particles with a radius smaller than 2 \textmu m. Additionally, a trace of solid phase CO$_2$ was identified in the bright band's material, which was not seen in the other spectra. These findings suggest that the material could be fresh and minimally affected by space-weathering. \cite{Dhingra2017SpatiallyVIMS} and the analysis in \autoref{sec:ices} show that the plumes consist mainly of crystalline water ice while \cite{Brown2006CompositionSurface} and \cite{Combe2019NatureEnceladus} detected CO$_2$ trapped in ice on the surface of Enceladus near the Tiger Stripes. However, the spectra of the plume and bright band's material in \autoref{fig:spec} remain different, leaving questions about the connection between the bright band's material and the fresh plume material. If the bright band's material is similar to fresh plume material, the plume spectrum would be expected to exhibit the CO$_2$ absorption feature as well. However, the plume's signal is noisier compared to the bright band at these wavelengths (as shown on the bottom left in \autoref{fig:spec}), which may obscure the CO$_2$ absorption in the plume spectrum. We noted that other spectral differences could be attributed to particle size distributions favouring larger particles in the bright band compared to the E ring and the plume. If the bright band indeed consists of fresh plume ice, why are its particles larger? Three possible explanations arise: (1) the bright band ice is more shielded from space weathering than ice sampled directly from the plume, (2) crystal growth occurs within the E ring, or (3) a process selectively removes the smallest particles in the plumes, leaving behind only the larger ones.

Lastly, the stripe features appear stationary when projected to an infinite distance in the J2000 reference frame. However, they are actually located in front of Enceladus, meaning they are relatively close to Cassini. The repeated observation of the bright band during a single flyby suggests a displacing phenomenon, which could be either a physically moving structure or an optical illusion in a homogeneous medium. One possible explanation is that the interaction between the apparent motion of the features along Cassini's line-of-sight and the reflection of light creates this effect, though the underlying mechanism remains poorly understood.

\begin{figure}[h]
    \centering
    \includegraphics[width=\linewidth]{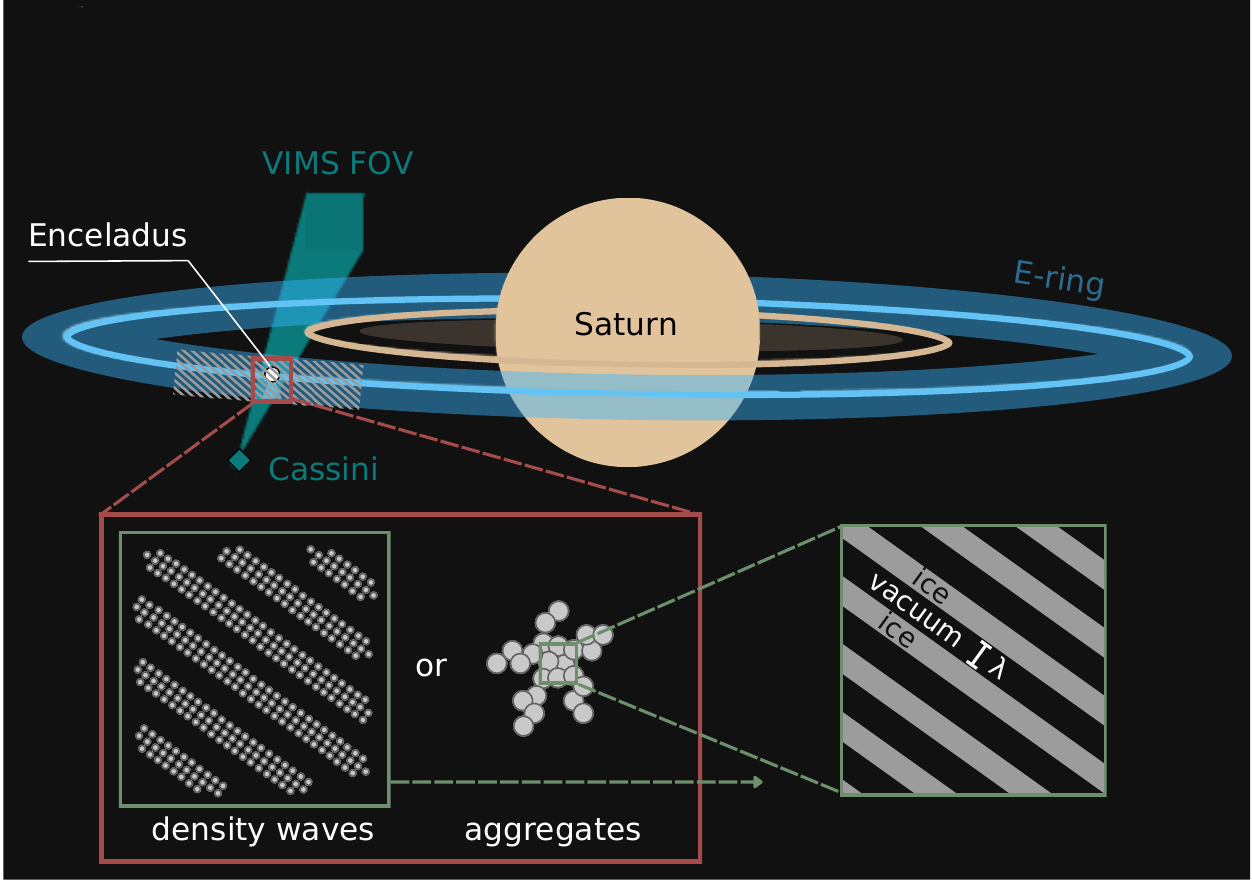}
    \caption{Visualisation of the hypothesis for the observed stripe features with either density waves or ice aggregates located at the E ring. Periodic structures in these features act as a reflection grating with groove spacing $\lambda$ (we suggest mm-scale) as shown by the zoomed-in view in the bottom right corner.}
    \label{fig:hypoview}
\end{figure}

\textit{\textbf{Hypothesis}: A millimetre-sized inclined ordered structure within the E ring.} The stripes are produced by an organised structure (acting as a reflection grating) positioned between Cassini and Enceladus; as for the stripes to appear in front of Enceladus, they must be produced by particles located physically in front of the moon. For flybys E17 and E19, this structure lies specifically between the orbits of Enceladus and Rhea, and for E13, between the G ring and Enceladus' orbit. This range of radial distances from Saturn spans the E ring's inner and outer regions. The fact that the stripe features are visible only when observing Enceladus and occur both interior and exterior to its orbit suggests that Enceladus is likely the source of the stripe phenomenon. A visual representation of the periodic E ring structure is given in \autoref{fig:hypoview}. We identify two potential processes within the E ring that could give rise to the formation of the observed periodic structure.

\textit{1) Periodic dust density waves:} E ring particles are organised into alternating high-density (more opaque) and low-density (more transparent) regions, due to external processes. \cite{Mitchell2015TrackingRing} found high-density sinusoidal structures within the E ring near Enceladus called 'tendrils'. These tendrils originate from the highest velocity particles ejected from the plumes and follow horseshoe orbits from Enceladus' leading and trailing edge. These horseshoe orbits' looping motion could potentially produce the necessary inclination and periodic structure for the stripe features. Simulations from \cite{Weiss2013TheRing} concluded that for particles smaller than $\sim$1 micron, the influence of Saturn's magnetic field blurs out the tendrils while larger particles $\sim$3 microns create distinct tendrils. This preference for larger particles aligns with the spectra shown in \autoref{fig:spec}. \cite{Hedman2021EvidenceRing} also demonstrated that similar high-density wave structures could result from rebound wakes, generated by gravitational perturbations on existing E ring particles due to Enceladus’ gravity.

\cite{Pickett2015} detected electrostatic solitary waves (ESWs) in data from Cassini's Radio and Plasma Wave Science Wideband Receiver (WBR) attributed to plasma activity within Saturn’s magnetosphere. These ESWs were observed inside 10 $\mathrm{R_S}$, particularly near Enceladus between 4 and 6 $\mathrm{R_S}$—the same region where stripe-like features were observed during flyby E17. ESWs can drive dust acoustic waves (DAWs) in the E ring’s dusty plasma, producing alternating high- and low-density regions that may appear as the observed chromatic stripes. DAWs have been modelled near Enceladus by \mbox{\cite{Singh2022}}, with predicted spatial spacings down to 11 meters. Laboratory experiments presented by \mbox{\citet{Merlino2009Dust}}, though not specific to Saturn’s environment, demonstrate that dust acoustic waves can form at millimetre wavelengths in the presence of micron-sized dust particles—consistent with our observational constraints. This supports the physical plausibility of ultrasonic-scale DAWs forming within Saturn’s magnetosphere. Cassini’s WBR may have missed these fine-scale structures due to its limited sampling frequency, which, based on our rough estimate, resolves features no smaller than a few centimetres.

\textit{2) Periodic ice grain structure:} beyond ice grains being periodically arranged, diffraction effects may also result from the shape of the ice crystals themselves, similar to the behaviour observed in photonic crystals \mbox{\citep{Joannopoulos1997PhotonicCrystals}}. In this case, icy aggregates in the E ring could create the required millimetre-scale periodicity. For example, \mbox{\cite{Gao2016AggregateEnceladus}} proposed that aggregates could form within Enceladus' plumes. It is also possible that the high-density waves mentioned earlier might promote particle aggregation. Yet, the optical behaviour of such photonic crystal-like aggregates is complex and highly sensitive to their structure, making it difficult to predict their signatures or directly relate them to the observed diffraction features.

\section{Conclusions}
\label{sec:conclusion}

\noindent In this article, we report the observation of stripe features simultaneously captured by Cassini’s ISS and VIMS-IR instruments during the Enceladus-targeted flyby E17. The stripes appear continuous across successive images when projected on the sky and are consistent between both instruments, excluding a stray light contribution and suggesting an external origin. Similar patterns were also observed in VIMS-IR data from flybys E13 and E19. A central bright band, most prominent at a wavelength of $\sim$5 microns, is believed to result from reflected light, with the additional stripes forming as higher-order diffraction peaks, much like a reflection grating.

We suggest the existence of an ordered structure within the E ring acting as a reflection grating with a millimetre-scale groove spacing. The structure, constrained between Saturn's G ring and Rhea's orbit, likely consists of fresh ice particles supplied by Enceladus' plumes. However, questions persist regarding the higher concentration of larger particles in the diffracting structure compared to those in the plumes. In addition, the stripe inclination of 43$^\circ$ to Saturn's ring plane is unusual for particles within the E ring. Potential explanations for the observed chromatic stripes include periodic enhancements in the E ring’s particle density—possibly caused by structures such as tendrils, wakes, or dust acoustic waves. Alternatively, periodic aggregates could also produce the required diffraction effects.

Recommendations for future research include filtering the VIMS cube to enhance the stripe contrast and reduce noise. This will aid in producing clearer spectral kymographs and deconstructing the contributions of the groove spacings and widths of the grating structure.

A specific configuration should be identified where the stripes remain fixed when projected at an infinite distance while physically situated between Cassini and Enceladus. This could be explored through ray tracing software or by setting up an experimental model replicating the scene's motions and illumination. 

The origin of the diffraction effects can be further investigated through diffraction analyses of ice crystal shapes and aggregates, as well as simulations of crystal orientations, vertical tendril structures, wakes, and dust acoustic waves within the E ring.

Further, fitting the VIMS spectra of the bright band to particle size distributions using Mie theory, as done by \cite{Hedman2009SpectralCassini-vims}, is recommended. A similar comparison with different ice mixtures could provide additional insights.

A survey of ISS and VIMS-IR data was conducted to find stripe features similar to those observed during flybys E13, E17, and E19. Expanding this survey to include VIMS-IR non-targeted flybys at high phase angles, primarily when Cassini is located between the G ring and Rhea's orbit, is advised. Investigating other instruments with similar near-infrared sensitivity and observation geometries would also be valuable.

\printcredits

\bibliographystyle{cas-model2-names.bst}

\bibliography{main}

\end{document}

\title{\textbf{Supplementary Material} \\ Peculiar Rainbows in Saturn’s E Ring: Uncovering Luminous Bands Near Enceladus}  
\author{Niels Rubbrecht$^{a}$, Stéphanie Cazaux$^{a}$, Benoît Seignovert$^{b}$, Matthew Kenworthy$^{a,c}$, \\ Nicholas Kutsop$^{d}$, Stéphane Le Mouélic$^{e}$, Jérôme Loicq$^{a}$}  
\date{}  

\maketitle  

\noindent\textit{{\small  
 $^{a}$ Faculty of Aerospace Engineering, Delft University of Technology, Kluyverweg 1, Delft, 2629 HS, The Netherlands \\  
$^{b}$ Observatoire des Sciences de l’Univers Nantes Atlantique (Osuna), CNRS UAR-3281, Nantes Université, France \\  
$^{c}$ Leiden Observatory, Leiden University, P.O. Box 9513, Leiden, 2300 RA, The Netherlands \\  
$^{d}$ Department of Astronomy, Cornell University, 404 Space Sciences Building, Ithaca, NY, 14853, United States \\  
$^{e}$ Laboratoire de Planétologie et Géosciences (LPG), CNRS UMR 6112, Nantes Université, Univ Angers, Le Mans Université, Nantes, 44000, France  
}}

\section{Analysis of Stripes Similar to Those in the E17 Flyby}  

This section presents the analysis of stripe features observed during Cassini’s Enceladus-targeted flybys E13 and E19, following a similar approach to the E17 analysis described in the paper.

\subsection{E13 Flyby}

\noindent \autoref{fig:4A_E13} presents a mosaic of the E13 flyby cubes at various wavelengths, revealing several notable similarities to the flyby E17 observations. Stripes are visible for wavelengths between 2.24 \textmu m and 4.26 \textmu m, and they remain continuous across the cubes, as can be seen in the second and third cubes at $\lambda = 3.10$ \textmu m. The stripes appear parallel and are inclined $16^\circ$ to the ecliptic, the same inclination of the stripes in flyby E17. The stripes in flyby E13 have an angular thickness of approximately 0.7--2.6 mrad, similar to the stripes observed in flyby E17.

Despite these similarities, several key differences exist between the E13 and E17 flybys. Unlike E17, no localised bright area is visible in E13, and the consecutive cube observations alternate between darker and brighter overall intensity. The stripes at $\lambda = 4.26$ \textmu m are significantly less bright than those at $\lambda = 3.10$ \textmu m, while for E17, there is no such difference. Additionally, E13 lacks the bright band and exhibits less global intensity overall at $\lambda =$ 5.11 \textmu m. Due to an instrument artefact creating horizontal stripes and an increased background brightness, ISS NAC images taken at the same time could not be brightened to detect faint stripe features in the plumes. 

Cassini's observer geometry compared to Saturn's ring plane is shown in \autoref{fig:7A_E13}. From the edge-on view, it can be seen that Cassini is located closer to the ring plane than in flyby E17, with a ring plane altitude ranging from 150 km to 225 km. From a top-down perspective, Cassini observes Enceladus from the inner edge of the E ring rather than the outer-edge as for flyby E17. The phase has increased by 4$^\circ$ compared to the maximum phase in flyby E17 with a phase angle of 164$^\circ$ at the start of the E13 approach to 165$^\circ$ at the end.

\subsection{E19 Flyby}

\begin{figure*}[h]
    \centering
    \begin{subfigure}[b]{0.44\textwidth}
        \centering
        \includegraphics[width=\textwidth]{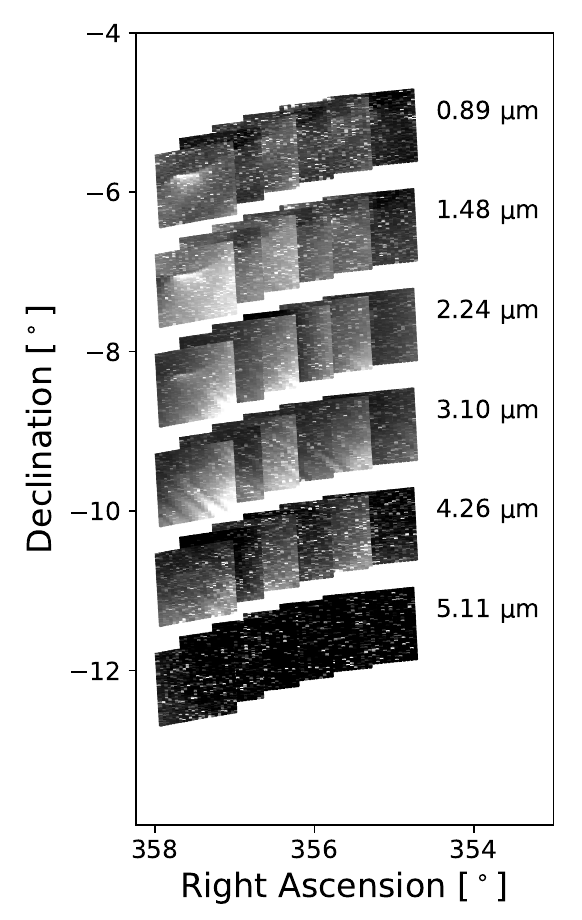}
        \caption{}
        \label{fig:4A_E13}
    \end{subfigure}
    \hspace{0.02\textwidth}  
    \begin{subfigure}[b]{0.44\textwidth}
        \centering
        \includegraphics[width=\textwidth]{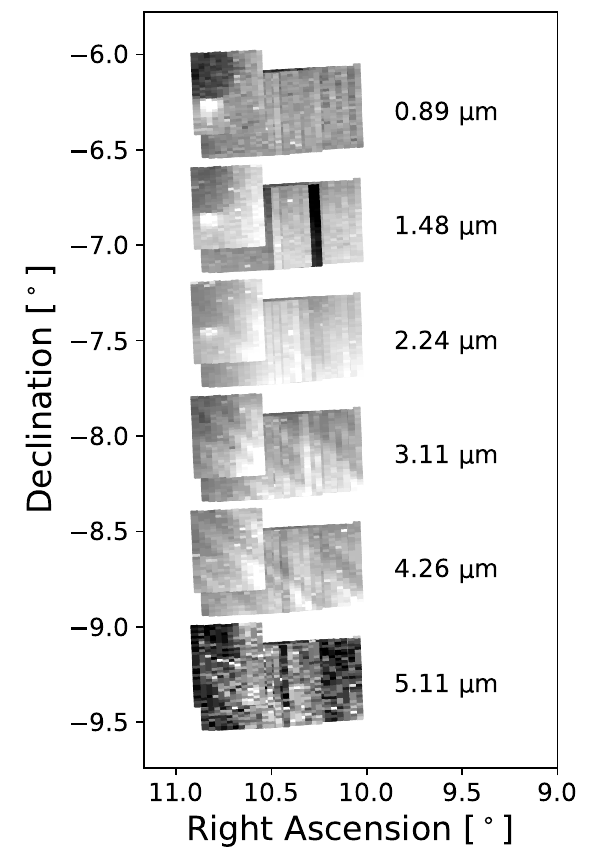}
        \caption{}
        \label{fig:4A_E19}
    \end{subfigure}
    \caption{Cassini flyby E13 (a) and E19 (b) VIMS mosaic in terms of right ascension and declination in the J2000 reference frame for a discrete set of wavelengths. The original mosaic is given on top; every following wavelength mosaic is shown with an additional declination for clarity.}
\end{figure*}

\noindent \autoref{fig:4A_E19} presents a mosaic of 11 VIMS cubes in flyby E19 at the same discrete wavelengths as the previous mosaics. All cubes in this range appear to be located within a bright area, with an intensity roughly consistent with the bright area observed in E17. Stripes are visible for the same spectral range as flyby E17 between 2.24 \textmu m and 4.26 \textmu m, and remain continuous across the cubes. The stripes are parallel and inclined $16^\circ$ to the ecliptic, just like the flyby E13 and E17 stripes. The stripe width is approximately 0.6--1.1 mrad.

Differences between E17 and E19 include a brightness increase in the low-phase corner of the E19 images that does not appear to correlate with the continuous stripes mapped in the RA and DEC projection. This non-continuous shift in brightness between cubes can be seen in the final cube, which has a slight DEC increase. Additionally, the E19 flyby lacks the bright band observed in E17. ISS NAC images at the same observation time during flyby E19 are available. However, no stripes could be detected. A potential cause for this could be that the ISS NAC stripe observations are phase-dependent and only show stripes for the highest phase images ($>$160$^\circ$) as seen in flyby E17. The flyby E19 ISS NAC images have a mean phase of 159.5$^\circ$ which is similar to the right-most images in the ISS E17 mosaic (Figure 2 from the paper). These images also do not show stripes. So, the observations between flyby E17 and E19 can be consistent, assuming a phase-dependent stripe visibility.

\begin{figure}[H]
    \centering
    \begin{minipage}{\linewidth}
        \centering
        \includegraphics[width=0.9\linewidth]{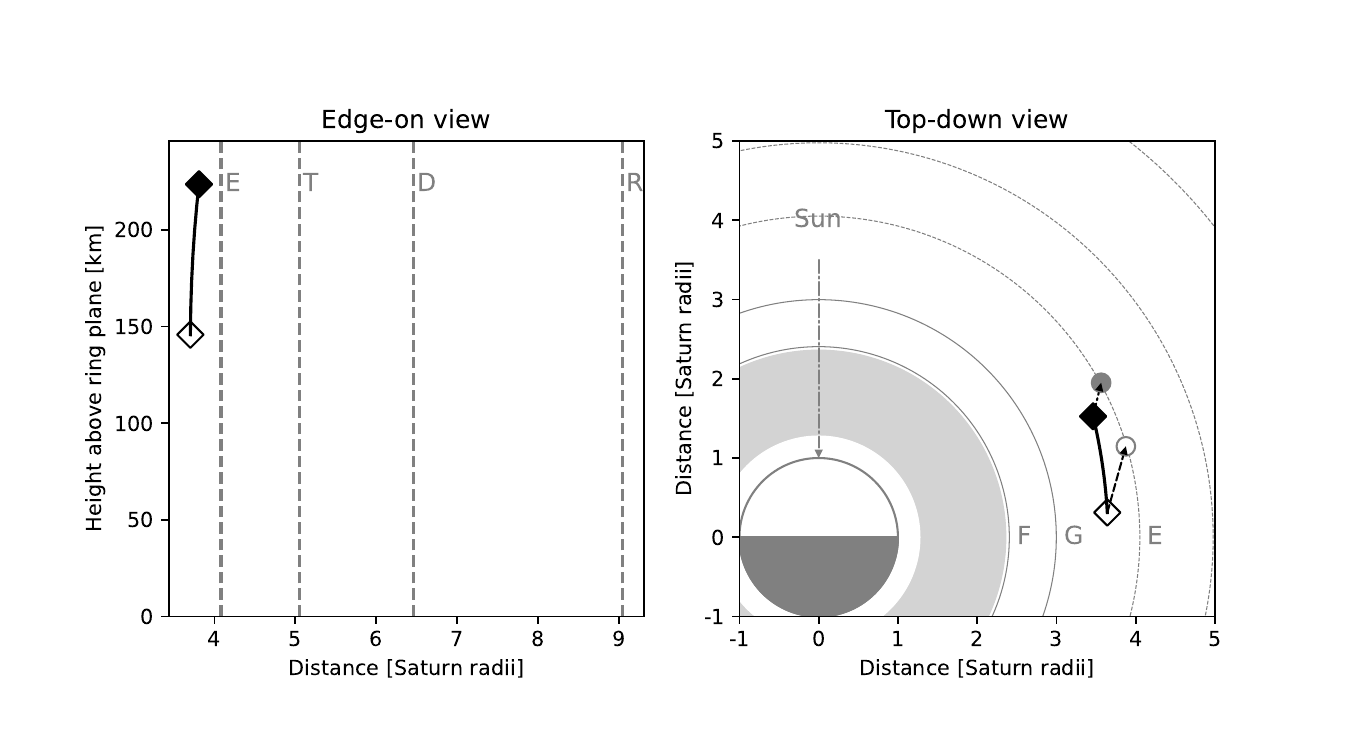}
        \caption{Cassini's E13 flyby geometry from 20/12/2010 22:53:39 to 21/12/2010 00:00:40 UTC. Cassini's position is marked by a diamond, and the moons' positions are indicated by circles. The open symbol represents the start position, and the filled symbol represents the end position. The orbits of Enceladus, Tethys, Dione, and Rhea (E, T, D, R) are represented by dashed grey lines, while the F-ring and G-ring (F, G) are shown with solid grey lines. The solar direction is indicated by a dash-dotted arrow, and the ISS/VIMS viewing direction by black dashed arrows.}
        \label{fig:7A_E13}
    \end{minipage}
    
    \vspace{1em} 
    
    \begin{minipage}{\linewidth}
        \centering
        \includegraphics[width=0.9\linewidth]{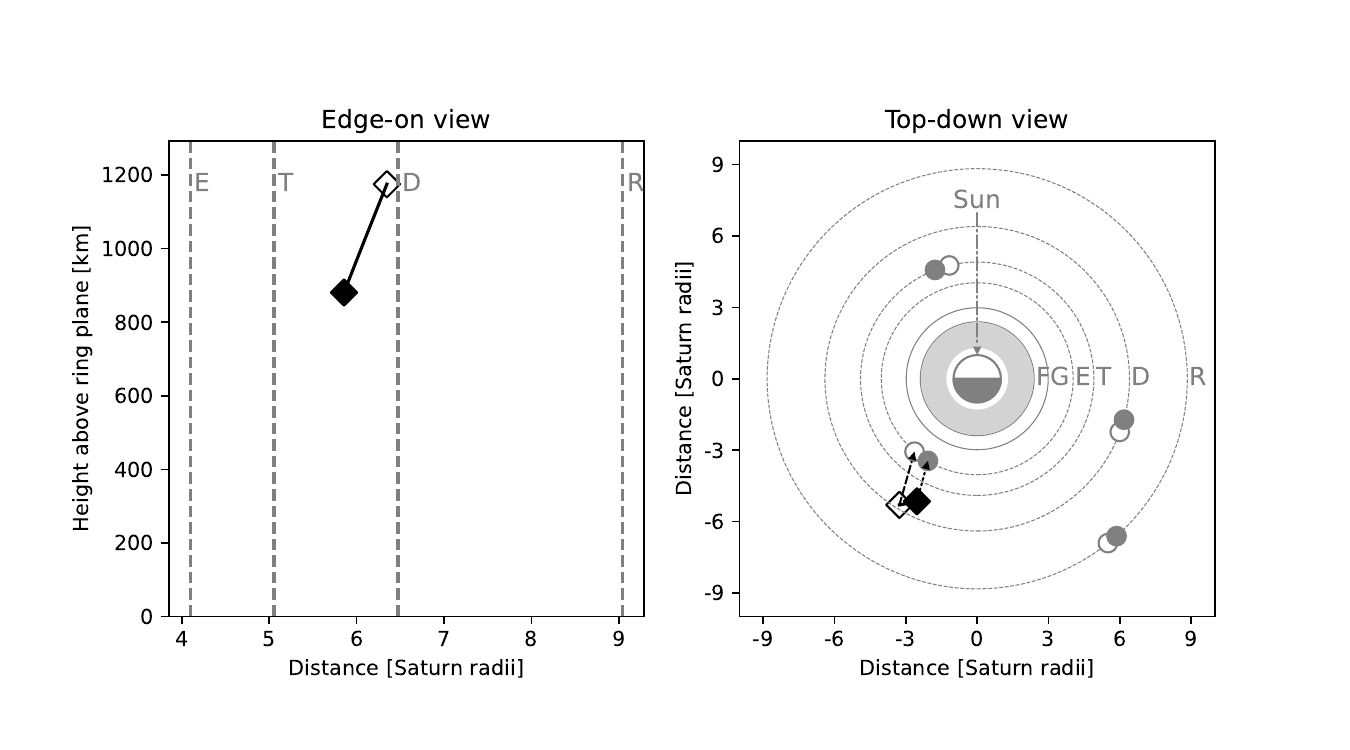}
        \caption{Cassini's E19 flyby geometry on 02/05/2012 from 05:07:13 to 06:02:00 UTC. Cassini's position is marked by a diamond, and the moons' positions are indicated by circles. The open symbol represents the start position, and the filled symbol represents the end position. The orbits of Enceladus, Tethys, Dione, and Rhea (E, T, D, R) are represented by dashed grey lines, while the F-ring and G-ring (F, G) are shown with solid grey lines. The solar direction is indicated by a dash-dotted arrow, and the ISS/VIMS viewing direction by black dashed arrows.}
        \label{fig:7A_E19}
    \end{minipage}
\end{figure}

\autoref{fig:7A_E19} illustrates a similar observation geometry in flyby E19 as in flyby E17, with Cassini positioned at the same altitude above the ring plane as during the bright area observation of E17. The radial distances from Saturn are also comparable, with the approach beginning shortly after crossing Dione’s orbit and concluding midway between the orbits of Tethys and Dione, mirroring the conditions during the bright area observations in E17 

\newpage
\section{Other Stripe Features}  

\begin{figure}[h]
    \centering
    \includegraphics[width=0.75\linewidth]{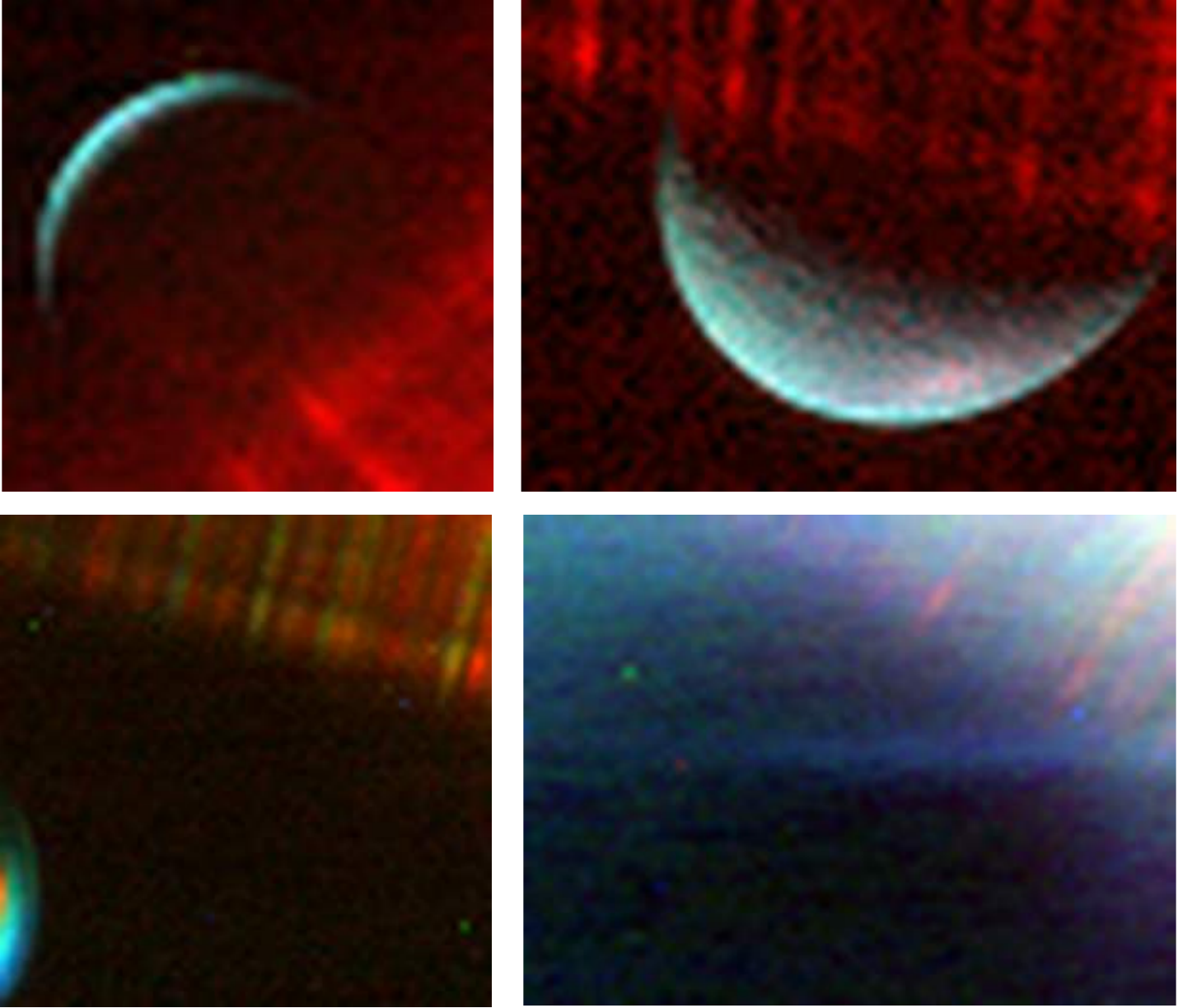}
    \caption{False colour image previews from the VIMS Data Portal \citep{LeMouelic2019TheMaps} of other stripe observations in the VIMS-IR channel. These show RGB intensities (3.1, 2.0, 1.78 \textmu m) of Dione cube 1818542624\_1 (top left), Rhea cube 1710080519\_1 (top right), and Tethys cube 1805596800\_1 (bottom right). Titan cube 1536295561\_1 (bottom left) is shown with RGB wavelengths of 2.77, 3.27, 3.32 \textmu m, respectively.}
    \label{fig:other_stripes}
\end{figure}

\autoref{fig:other_stripes} presents stripe observations that could not be linked to those observed in flybys E13, E17, and E19.

\section{Spectral Kymographs E13 and E19 Flyby}

\begin{figure}[H]
\centering
\begin{minipage}{.45\textwidth} 
    \centering
    \includegraphics[width=\linewidth]{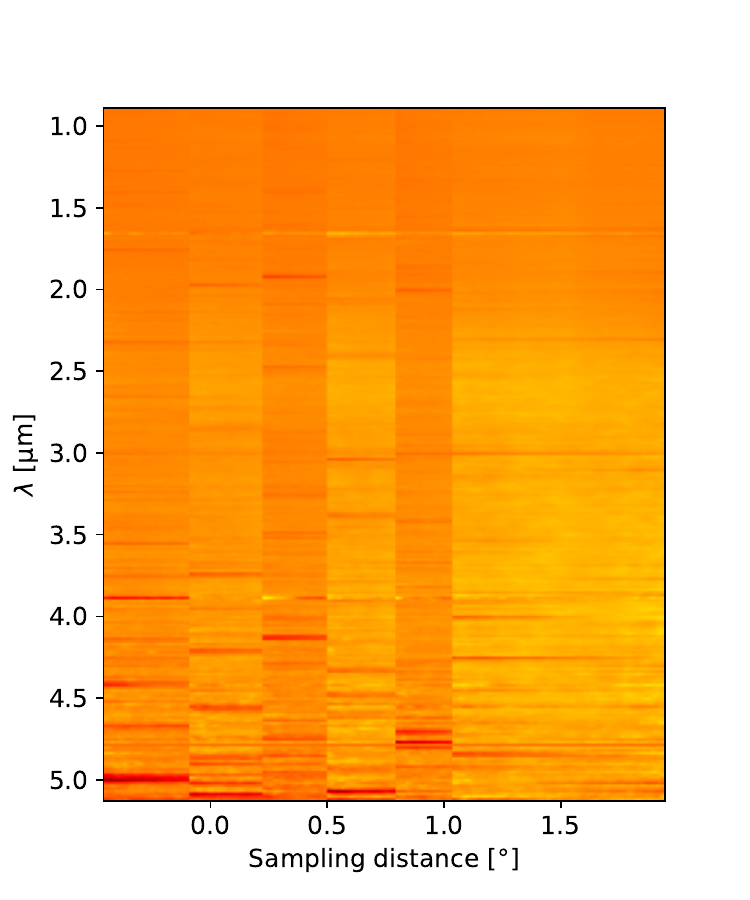}
    \caption{Spectral kymograph mosaic of Cassini flyby E13.}
    \label{fig:kymo_mos_e13}
\end{minipage}
\hspace{0.05\textwidth} 
\begin{minipage}{.45\textwidth}
    \centering
    \includegraphics[width=\linewidth]{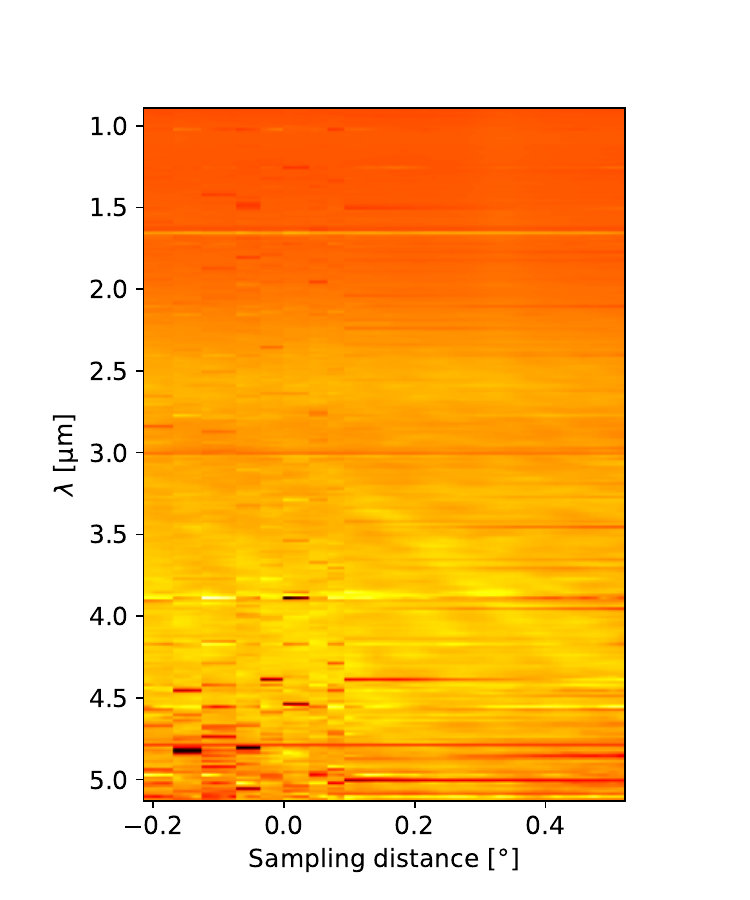}
    \caption{Spectral kymograph mosaic of Cassini flyby E19.}
    \label{fig:kymo_mos_e19}
\end{minipage}
\end{figure}

Spectral kymographs for the flyby E13 and E19 VIMS mosaics (\autoref{fig:4A_E13} and \autoref{fig:4A_E19}) are shown in \autoref{fig:kymo_mos_e13} and \autoref{fig:kymo_mos_e19} respectively. No vertical or slanted lines can be seen in flyby E13. However, E19 does display at least two distinct slanted lines. No vertical line representing the bright band can be seen in the E19 kymograph, which matches the absence of the bright band in the E19 VIMS mosaic in \autoref{fig:4A_E19}. From \autoref{fig:7A_E19}, the E19 observations occur at a location similar to E17 just after crossing Dione's orbit. In flyby E17, it was reported that the bright band was observed in flyby E17 just before this crossing. Therefore, assuming the bright band remains in the same radial distance to Saturn, flyby E19 VIMS observations should occur just after the bright band is observable. The earliest observations in the kymographs are on the left, while the latest are on the right. In the E17 spectral kymograph, lines to the right of the bright band observation have a negative slope. Since the flyby E19 cubes are assumed to occur after the bright band observation, the kymograph lines for flyby E19 should also have this negative slope. This is true from observing \autoref{fig:kymo_mos_e19}. 

\bibliographystyle{cas-model2-names-v2.bst}
\bibliography{sup.bib}